# An untrained deep learning method for reconstructing dynamic magnetic resonance images from accelerated model-based data

Kalina P. Slavkova[1], Julie C. DiCarlo[2,3], Viraj Wadhwa[4], Sidharth Kumar[4], Chengyue Wu[2], John Virostko[2,3,6,7], Thomas E. Yankeelov[2,3,5,6,7], Jonathan I. Tamir[2,4,6*]

[1] Department of Physics, The University of Texas at Austin, Austin, USA

[2] The Oden Institute for Computational Engineering and Sciences, The University of Texas at Austin, Austin, USA

[3] Livestrong Cancer Institutes, The University of Texas at Austin, Austin, USA

[4] Department of Electrical and Computer Engineering, The University of Texas at Austin, Austin, USA

[5] Department of Biomedical Engineering, The University of Texas at Austin, Austin, USA

[6] Department of Diagnostic Medicine, The University of Texas at Austin, Austin, USA

[7] Department of Oncology, The University of Texas at Austin, Austin, USA

**\*** Please address correspondence to**:**

Jonathan I. Tamir, Ph.D.

The University of Texas at Austin

Department of Electrical and Computer Engineering

EER 7.872

2501 Speedway

Austin, TX 78712, USA

E-mail: jtamir@utexas.edu

**ABSTRACT**

**Purpose:** To implement physics-based regularization as a stopping condition in tuning an untrained deep neural network for reconstructing MR images from accelerated data.

**Methods:** The ConvDecoder neural network was trained with a physics-based regularization term incorporating the spoiled gradient echo equation that describes variable-flip angle (VFA) data. Fully-sampled VFA *k*-space data were retrospectively accelerated by factors of $R=\{8,12,18,36\}$ and reconstructed with ConvDecoder (CD), ConvDecoder with the proposed regularization (CD+r), locally low-rank (LR) reconstruction, and compressed sensing with L1-wavelet regularization (L1). Final images from CD+r training were evaluated at the '*argmin*' of the regularization loss; whereas the CD, LR, and L1 reconstructions were chosen optimally based on ground truth data. The performance measures used were the normalized root-mean square error, the concordance correlation coefficient (CCC), and the structural similarity index (SSIM).

**Results:** The CD+r reconstructions, chosen using the stopping condition, yielded SSIMs that were similar to the CD ($p=0.47$) and LR SSIMs ($p=0.95$) across *R* and that were significantly higher than the L1 SSIMs ($p=0.04$). The CCC values for the CD+r $T_1$ maps across all *R* and subjects were greater than those corresponding to the L1 ($p=0.15$) and LR ($p=0.13$) $T_1$ maps, respectively. For $R \geq 12$ ($\leq 4.2$ minutes scan time), L1 and LR $T_1$ maps exhibit a loss of spatially refined details compared to CD+r.

**Conclusion:** The use of an untrained neural network together with a physics-based regularization loss shows promise as a measure for determining the optimal stopping point in training without relying on fully-sampled ground truth data.

# 1 INTRODUCTION

MRI is known to have superior tissue contrast compared to other imaging modalities, such as computed tomography (CT)[1]. In imaging ischemic stroke patients, for instance, MRI was shown to be substantially more sensitive in detecting lesions (93.5% positivity rate[2]) compared to CT (85% positivity rate[2])[3]. In breast imaging[4], for example, MRI has shown greater sensitivity than mammography with recent advancements[5–7]. While sensitivity is superior, MRI may lack specificity in some cases compared to other imaging techniques[4]. False positives in general may lead to unnecessary biopsies with concomitant stress to the patient and cost to the healthcare system[6]. Thus, there is substantial effort toward increasing the diagnostic specificity of MRI through the development of quantitative imaging schemes that provide numeric data in addition to standard of care (SOC) images[8,9] for informing patient care.

Quantitative imaging involves fitting MRI data to a known physical model to estimate parameters that represent tissue characteristics[10]. For instance, the fast spoiled gradient echo (SPGR) sequence equation is fit to data from a $T_1$-weighted variable-flip angle (VFA) acquisition to estimate the tissue $T_1$ values at each voxel[11], where $T_1$ values have been shown to distinguish some pathologies[12]. $T_1$ parameter mapping is also an important component in other quantitative schemes, such as pharmacokinetic analysis of dynamic contrast-enhanced (DCE) MRI in breast cancer imaging[13] and brain imaging[12,13,14] for diagnosis[13,15] and treatment monitoring[16–19].

Despite the utility of quantitative imaging, scan time constraints remain a barrier to incorporating quantitative imaging schemes into SOC protocols. To combat this issue, many investigators are attempting to reduce scan time necessary for clinical application through model-based, multi-contrast methods. One such example is Magnetic Resonance Fingerprinting (MRF) for collecting multiparametric maps in a single, fast scan using dictionary matching[20] with unique, non-clinical sequences[21]. MR multitasking[22–24] is another multiparametric mapping technique that combines different imaging dynamics along multiple orthogonal dimensions that are subsequently resolved as a low-rank tensor[22], naturally extending low-rank matrix priors. Echo-planar time-resolved imaging (EPTI) is another promising method for fast quantitative imaging that involves optimally under-sampling *k*-space along both spatial and temporal dimensions without blurring artifacts from traditional echo-planar imaging[25,26]. These are rich multi-contrast acquisitions that incorporate a well-defined physical model with great potential for compatibility with deep learning advancements.

Model-based reconstruction[27] *via* trained deep learning methods[28–30] has shown promise in utilizing deep convolutional neural networks (CNNs) for reconstructing images with sufficient spatial and/or temporal resolution from highly accelerated raw MRI data for successful subsequent analysis with appropriate models. One such method is the Model-Based Deep Learning (MoDL)[28] framework by Aggarwal *et al.*, which introduces an end-to-end supervised neural network-based noise estimator as a regularization term. Studies relying on trained methods, however, are limited by the availability of sufficiently large, curated training datasets. This is an issue that self-supervised and untrained methods[31–35] circumvent by relaxing the need for fully-sampled ground truth data. The reference-free latent map extraction (RELAX) method[36] trains a U-Net[37] architecture in a self-supervised manner to map under-sampled data to quantitative arameter maps. While the RELAX method does incorporate physical dynamics, the method still requires a large training set. The Self-Supervision *via* Data Under-sampling (SSDU) method[30] is a self-supervised method that also relies only on a large training set of available under-sampled data that is split into separate sets *via* multiple masks.

Conversely, untrained generative neural networks – such as Deep Image Prior[35], DeepDecoder[32], and ConvDecoder[31] (where the term "untrained" was coined) – do not rely on a training data set. Instead, they operate on a single under-sampled input and are thus scan-specific, relying on the neural network structure for implicit regularization. The ConvDecoder has been shown to substantially outperform conventional methods based on sparsity and low-rank principles and has been shown to perform similarly to trained methods in some cases[31]. Still, these methods neither incorporate the known physical dynamics that further constrain the image reconstruction inverse problem nor include a stopping condition for determining when an optimal solution has been reached; therefore, they are prone to overfitting and can result in image artifacts[27,31,33,35]. Yaman *et al.*[38] and, recently, Leynes *et al.*[39] propose regularizations based on data-consistency and statistical properties, respectively, that serve as stopping conditions for scan-specific reconstructions, but these methods do not include the quantitative model.

To address the need for a physics-informed untrained method with a reliable stopping condition for high-resolution dynamic MRI data amendable to quantitative analysis, we previously applied the ConvDecoder (CD)[31] method with an SPGR-based regularization term (CD+r)[40] as a proof-of-principle for reconstructing multi-dimensional images from simulated accelerated VFA data. Here, we formalize the CD+r method and apply it to retrospectively accelerated 3D VFA data collected from three healthy subjects. We compare the image reconstruction quality and $T_1$ map output of fully-sampled VFA data to the output of retrospectively accelerated VFA data reconstructed with CD+r, CD, parallel imaging with locally low-rank regularization (LR)[41], and compressed sensing (L1)[42]. We show that the physics-based regularization term in training CD+r provides a natural stopping condition for reconstruction that is blind to ground truth data, yielding $T_1$ maps that have a higher agreement with reference $T_1$ maps compared to $T_1$ maps derived from the CD, L1, and LR methods with ground truth-informed optimized hyperparameters. CD+r is a promising untrained method that leverages MRI physics for accelerated multi-contrast data acquisition where there is no fully-sampled ground truth as a benchmark for tracking reconstruction performance. Additionally, CD+r is useful for imaging modalities and anatomies with insufficient training examples for supervised and self-supervised methods. While CD+r is evaluated for a specific case (i.e., $T_1$-mapping), it applies to any quantitative model-based MRI dataset with an appropriate model chosen as the regularization term, making it a useful physics-informed solution to model-based MR image reconstruction from under-sampled data.

## 2   METHODS

*2.1 Definition of the MRI forward operator*

We define a sampling mask, $M$, which is multiplied with fully sampled $k$-space, $y$, data to yield retrospectively under-sampled $k$-space data, $y_u$. The sampling mask is combined with the coil sensitivities[43], $S$, and the Fourier transform, $F$, to arrive at the MRI forward operator, $A$, defined as

$$A = MFS. \qquad [1]$$

For a mask of all ones, corresponding to fully sampling data, $A$ operator maps the image, $x$, to $k$-space samples, $y$, through the relationship $y = Ax$. In the case of accelerated data, $y_u$, the zero-filled image, $x_{adj}$, can be computed as $x_{adj} = A^H y_u$, where $A^H$ is the adjoint of $A$.

*2.2  Physics-based model definition and quantitative parameter mapping*

While the proposed method in this work is compatible with any quantitative model-based MRI dataset, VFA data were chosen for a proof-of-principle analysis because of the ease of acquiring fully-sampled raw data. Because a $T_1$ map from VFA data is an important component of other quantitative acquisitions, such as DCE-MRI, demonstrating a speedup with this method already presents a speedup of a quantitative image exam. For a VFA acquisition, the SPGR equation describes the behavior of the image signal intensity (*SI*) as a function of RF flip angle $\theta$. This equation is defined as follows[44]:

$$SI(\theta) = S_o \frac{\sin(\theta)(1-e^{-TR/T_1})}{(1-\cos(\theta)e^{-TR/T_1})}, \qquad [2]$$

where $T_1$ is the longitudinal relaxation time, *TR* is the repetition time, and $S_o$ is the equilibrium signal intensity that is proportional to the spin density. To fit Eq. [2] to signal intensity images, **x**, collected at multiple flip angles, we implemented dictionary matching[20] using 2,000 $T_1$ values linearly spaced between 50 ms and 4000 ms. The result is a $T_1$ map, **$T_1$**, corresponding to a simulated image defined as **$x_m$**, which is the set of VFA images outputted from fitting **x** to $SI(\theta)$ via dictionary matching (see Supplemental Methods on dictionary matching).

*2.3 Implementation of the proposed deep learning framework*

The ConvDecoder[31], **G**(*w*), is a generative CNN with weights *w* that maps a low-dimensional vector of random noise (Gaussian noise with mean 0 and standard deviation 1) to an image. It is an extension of the original DeepDecoder[32] framework with added convolutional operations, where the weights are iteratively optimized on a single dataset as opposed to trained on a collection of training data (i.e. supervised and self-supervised methods).

The cost function, *L*(**G**(*w*)), for training **G**(*w*) was defined as follows:

$$L(\mathbf{G}(w)) = \min_w \|y_u - A\mathbf{G}(w)\|_2^2. \qquad [3]$$

This cost function strictly enforces data consistency between the inputted accelerated *k*-space data, $y_u$, and the network output, **G**(*w*), passed through the forward operator, **A** (see Eq. [1]). Even though the problem is ill-posed, the CNN provides implicit regularization favoring natural images. The original implementation of the ConvDecoder (CD) architecture (see Supplemental Figure S1 for the network structure) was adapted into DeepInPy[45], a Python-based framework for training deep learning methods specific to MRI reconstruction problems. Each under-sampled dataset presented to the CD network is reconstructed separately and independently. The network was calibrated using the ISMRM/NIST phantom[46] scanned with the same equipment and protocol described below in Section 2.4 (see Supplemental Methods for phantom details).

The weights were updated through backpropagation using the Adam optimizer[47], where a step size of $\delta = 0.01$ was empirically chosen based on calibration with the phantom scans. The network was trained for 10,000 training steps. The optimal reconstruction (complex-valued, coil-combined VFA image series) for each CD experiment was chosen by evaluating the *argmin* of the final normalized root-mean square error (NRMSE) curve as a function of training steps as a "best-case" reconstruction compared to the ground truth. The NRMSE was retrospectively smoothed with the Savitzky-Golay method[48] prior to finding the *argmin* (Supplemental Figure S2).

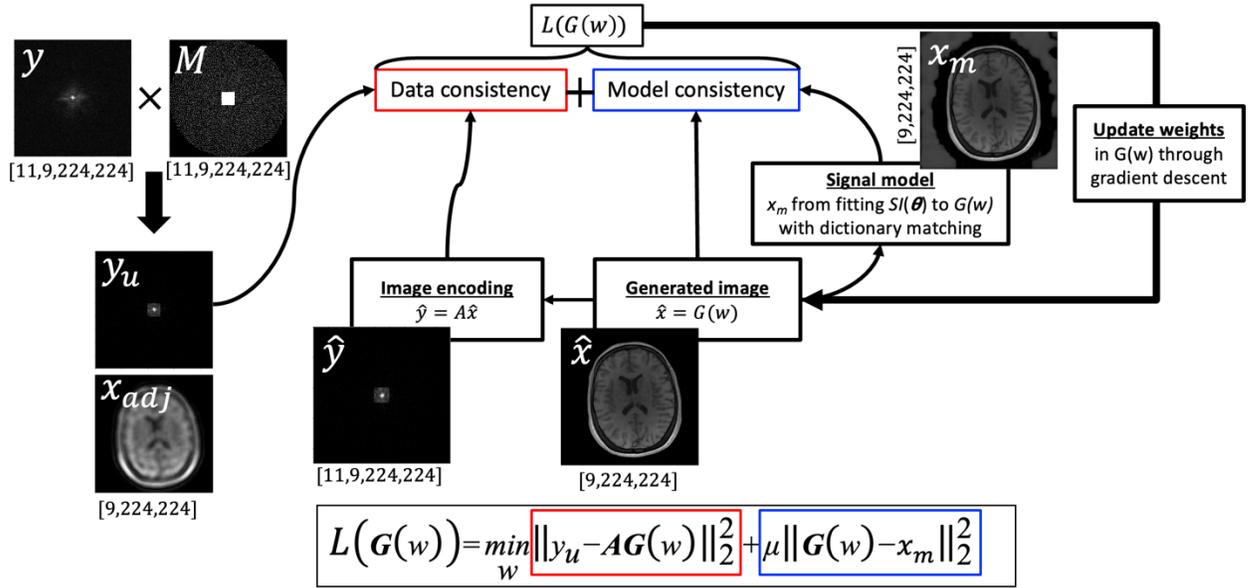

**Figure 1: Overview of the CD+r reconstruction pipeline.** A fully sampled *k*-space dataset, *y* (where the shape is defined below the labelled panel), is multiplied with a Poisson-disc sampling mask (*M*) to arrive at under-sampled k-space, $y_u$ (the corresponding zero filled image, $x_{adj}$, is shown below it). The ConvDecoder takes noise as input and outputs an estimated VFA image series as $\hat{x}=G(w)$. Using this estimate, a second VFA image series $x_m$ is computed by dictionary matching the output of $\hat{x}$ fitted to the signal model, $SI(\theta)$. These two images are inputted into the model consistency term of the loss function at each training step, which compares $y_u$ to $A\hat{x}$ and $G(w)$ to $x_m$. The weights of $G(w)$ are updated through gradient descent based on the loss $L(G(w))$, and a new $\hat{x}$ is generated. As described in the main text, $x_m$ is updated every five training steps. The size of all k-space data is reported as [# of coils, # of flip angles, X, Y], and the size of all images is reported as [# of flip angles, X, Y].

For ConvDecoder with physics-based regularization (CD+r), the CD network structure (Figure S1) and the same scheme were used; however, the cost function in Eq. [3] was altered to reflect the regularization term:

$$L(G(w)) = \min_{w} \|y_u - AG(w)\|_2^2 + \mu \|G(w) - x_m\|_2^2, \quad [4]$$

where $\mu$ is the regularization parameter, and $x_m$ is the SPGR model output computed through dictionary matching as described in Section 2.2. The proposed pipeline for CD+r is visually outlined in Figure 1. The model, $x_m$, was updated from $G(w)$ every $j=5$ training steps through alternating minimization[49], where the value of $j$ was empirically optimized (see Supplemental Methods for more detail). Four values of the regularization parameter were selected after investigating many values across scales during hyperparameter optimization: $\mu \in \{0.05, 0.10, 0.50, 1.0\}$. The regularization loss was retrospectively smoothed with the Savitzky-Golay method[48] (Supplemental Figure S2), and the optimal reconstruction for each experiment was selected by computing the *argmin* of the regularization loss (noting that CD+r is blind to the ground truth data).

Reconstruction-based $T_1$ maps corresponding to the optimal reconstructions outputted from CD and CD+r were a biproduct of dictionary matching in solving for $x_m$ during optimization[20]. All deep learning experiments were performed on NVIDIA V100 GPUs with 16 GB memory. Code to reproduce our results is available at https://github.com/kslav/cdr_mri.git.

*2.4 Data acquisition*

Three healthy subjects (denoted S1, S2, and S3) were scanned using a 3T MAGNETOM Vida MRI scanner equipped with a 32-channel head coil array (Siemens, Munich, Germany) under approval of the Institutional Review Board and with informed consent. Fully sampled VFA data from each subject were collected using a 3D SPGR sequence with an acquisition matrix of 224×224×224 over a (22.4 cm)³ field of view (i.e., 1 mm³ isotropic resolution), $TR/TE = 6.10/2.75$ ms, and nine flip angles from 4° through 20° and linearly spaced in 2° increments. No parallel imaging was employed so that all *k*-space samples were collected, which required 50 minutes of scanning.

A representative dynamic slice (as a function of $\theta$) from each dataset was selected for analysis following an inverse Fourier transform along the read-out direction. The magnitudes of the VFA anatomical images at all nine flip angles are displayed in Supplemental Figure S3 (images and data are complex-valued, see Supplemental Methods for more information on data formatting). To eliminate correlated noise across the sensitivity coils, noise pre-whitening was performed on all three datasets using a noise pre-scan and the BART toolbox[50]. Finally, software SVD-based coil compression was applied to the raw *k*-space datasets[50–52] to shrink the coil space down to 11 signal-contributing virtual coils, thereby eliminating coils contributing less than 1% of the total signal (see Supplemental Figure S4).

## 2.5 Simulated data generation without model error

To determine how quantitative model error affects the performance of the proposed physics-regularized method, datasets with no model error were first created from the raw data by synthesizing the forward equation. A set of three reference $T_1$ maps and modelled images were estimated using dictionary matching applied to the three fully sampled datasets. Each resultant simulated VFA image was multiplied by coil sensitivity maps (described in Section 2.6 below) to arrive at simulated coil images. The *k*-space data were subsequently computed by taking the Fourier transform of these coil images. Finally, noise was added to the *k*-space of each simulated image based on the native SNR of the raw data. The result was a set of three noisy, simulated datasets derived from the three raw datasets without VFA model error.

## 2.6 Preparation of retrospectively accelerated data

After acquiring and pre-processing the raw *k*-space and generating simulated data by the methods in Sections 2.4 and 2.5, respectively, the raw and simulated data were subsequently normalized to have a norm of 1,000 (see Supplemental Methods on phantom scans). The coil sensitivities, ***S***, were computed using the ESPIRiT algorithm[53] implemented in BART. The coil sensitivity maps corresponding to each raw dataset were used in Section 2.5 for the simulated data as the modelled anatomy remained the same.

Next, the sampling masks, ***M***, were created. A set of unique variable-density 2D Poisson-disc sampling patterns with fully sampled auto-calibration regions of size 25×25 was generated using BART and stored in a 3D volume of masks matching the VFA data dimensions. We chose acceleration factors of $R \in \{8, 12, 18, 36\}$, corresponding to scan durations of {6.25, 4.17, 2.78, 1.4} minutes; thus, a set of ***M*** was generated for each value of *R* (Figure S4). The three fully sampled raw and simulated datasets were all retrospectively accelerated by taking the product of *M* with the fully sampled *k*-space. An example set of corresponding zero-filled images is displayed in Supplemental Figure S5.

### 2.7 Implementation of the baseline methods of reconstruction

Compressed sensing (L1)[42,54] and locally low-rank reconstruction (LR)[41,55], both implemented using BART, were chosen as the baseline methods against which the performance of the proposed method was evaluated. These methods were chosen because they are commonly used, are effective, do not rely on training data, and lack a stopping condition independent of ground truth data.[31] For each set of raw and simulated data, the optimal reconstructed images from L1 and LR were found by optimizing over the regularization parameter, $\lambda$, and the number of iterations, $N$, (available in Table 1 for the raw data and Supplemental Table S1 for the simulations) through the minimization of the NRMSE computed between the ground truth images and the reconstructions. Reconstruction-based $T_1$ maps corresponding to the final L1 and LR-reconstructed images were computed using dictionary matching.

### 2.8 Performance measures and statistical analysis

The NRMSE, concordance correlation coefficient (CCC)[56], and structural similarity index (SSIM)[57] were computed to assess the performance of the proposed framework. We used the fully sampled raw *k*-space data as reference ground truth scans and computed the corresponding reference $T_1$ maps for all three subjects. This allowed us to quantitatively benchmark the performance of the reconstruction schemes (Sections 2.3 and 2.7) against the ground truth image quality. The NRMSE and SSIM were used to quantify the performance of the VFA image series, whereas the NRMSE and CCC were used to measure the agreement between the reference $T_1$ map (Section 2.5) and the reconstruction-based $T_1$ maps. The Wilcoxon rank sum test was used to compute *p*-values when comparing performance metrics across methods. Generally, average values are accompanied with "(± standard deviation)."

## 3 RESULTS

Across all methods, acceleration factors, and simulated and raw data, a total of 168 reconstruction experiments were performed. Each CD experiment took 3.16 (±0.24) hours to complete on average, and each CD+r experiments took 5.15 (±0.20) hours to complete.

*3.1 Applying the regularization loss as a stopping condition*

The suitability of the regularization loss as a stopping condition was evaluated across the CD+r experiments in the simulated and raw data settings at four values of $\mu$ and $R$ (Figure 2, Supplemental Figure S6). The NRMSEs, SSIMs, and optimal number of training steps can be found for these experiments for the raw data in Supplemental Table S2 and the simulated data in Supplemental Table S3. Bolded values in Supplemental Tables S2 and S3 indicate the highest performing metrics, which, in Supplemental Table S2, correspond predominantly to $\mu=0.1$ for $R \leq 18$ and to $\mu=1.0$ for $R=36$. In Supplemental Table S3, the highest performing metrics correspond to $\mu=0.5$ and $\mu=1.0$. Figure 2A displays training curves and performance measures for the simulated case for subject S3 at $R = 12$. In this figure, the smoothed regularization loss curves reached global minima at 1,981 and 2,600 training steps for $\mu = 0.05$ and $\mu = 0.10$, respectively, with NRMSEs of 0.13. These two results closely match the NRMSE of the optimal CD reconstruction ($\mu = 0$) (based on the global minima of the NRMSE curve), was 0.13. Similar NRMSE values were found for $\mu = 0.50$ and $\mu = 1.0$.

The added physics-informed regularization in Figure 2A yielded lower NRMSE values across all four $\mu$'s at later epochs (roughly post 3,000) compared to CD. At the final training step of 9,999 (Figure 2A), the CD reconstruction had a NRMSE of 0.28 compared to 0.20 for the CD+r experiment with the greatest final NRMSE, corresponding to $\mu = 0.05$. The CD+r reconstructions for subject S3 yielded an average NRMSE and SSIM across $\mu$ of 0.13 (±0.002) and 0.91 (±0.01), respectively, compared to the CD reconstruction with a NRMSE of 0.13 and a SSIM of 0.89. Averaging over all simulated subject datasets at $R=12$, the CD+r reconstructions yielded an average NRMSE and SSIM of 0.15 (±0.01) and 0.89 (±0.03), respectively, compared to CD reconstructions with an average NRMSE of 0.15 (±0.01) and an average SSIM of 0.88 (±0.03).

The raw data analysis for subject S3 at $R=12$ is displayed in Figure 2B. Larger regularization rates of $\mu = 0.50$ and $\mu = 1.0$ corresponded with a greater duration of training compared to the simulated results in Figure 2A. Like the NRMSE curves in Figure 2A, the added physics-informed regularization in Figure 2B lowered NRMSEs for all four $\mu$ values at later epochs compared to CD in Figure 2B. At the final training step, the CD reconstruction had a NRMSE of 0.32 compared to 0.19 for the worst-performing CD+r experiment with $\mu = 0.05$. The CD+r reconstructions for subject S3 yielded an average NRMSE and SSIM across $\mu$ of 0.12 (±0.005) and 0.91 (±0.01), respectively, compared to the CD reconstruction with a NRMSE of 0.12 and a SSIM of 0.91. Averaging over all raw subject datasets at $R=12$, the CD+r reconstructions yielded an average NRMSE and SSIM of 0.17 (±0.04) and 0.86 (±0.04), respectively, compared to CD reconstructions with an average NRMSE of 0.15 (±0.02) and an average SSIM of 0.88 (±0.02).

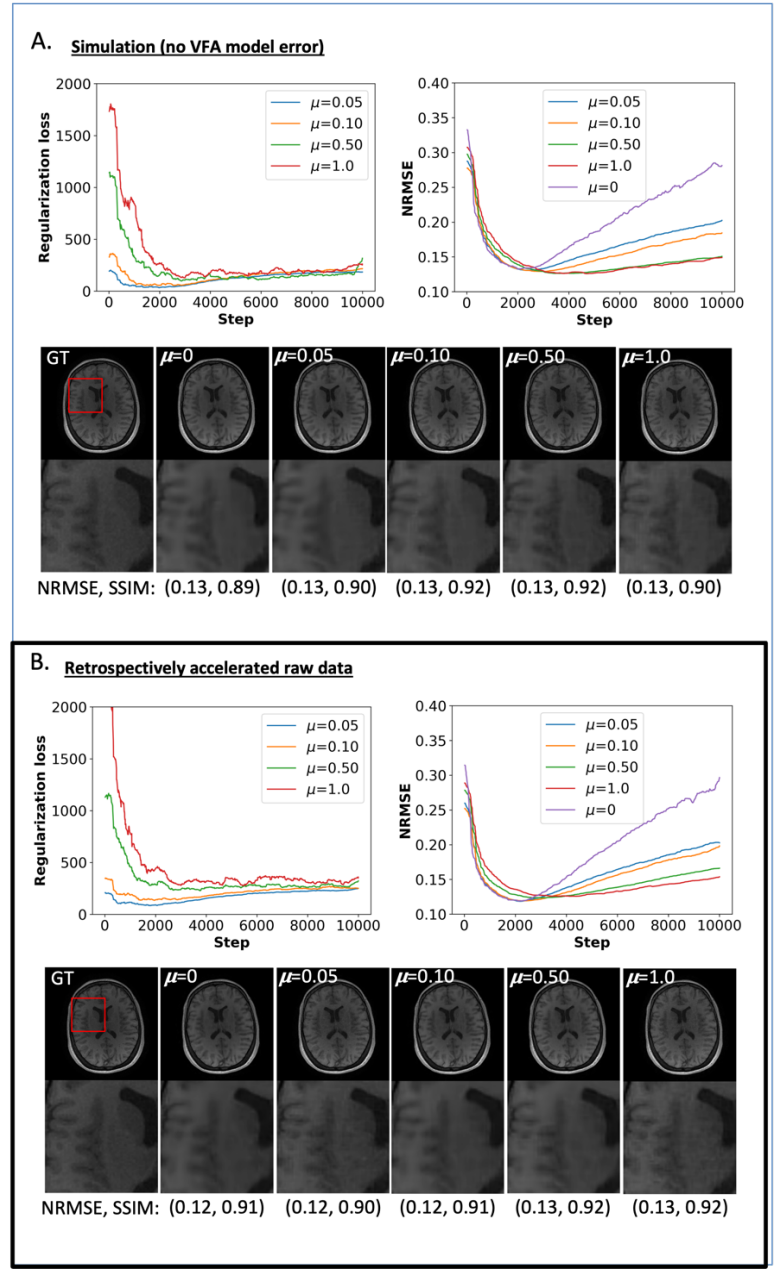

**Figure 2: Regularization loss and NRMSE in a representative dataset for $R = 12$.** A) The top left and top right plots show the regularization loss and the NRMSE, respectively, as a function of training steps for the simulated data. Below these plots, representative VFA reconstructions at $\theta=10°$ at four regularization rates, $\mu$, are juxtaposed with the ground truth (GT) image. The numbers in parentheses below the anatomical images inform the NRMSE and SSIM for the optimal reconstruction based on the *argmin* of the regularization loss (for non-zero $\mu$) and the NRMSE curve (for $\mu$=0). B) Analogous to the information in panel (A) for the raw data analysis. The results in both panels are similar in terms of NRMSE and SSIM; though, more training steps were required to reach the optimal stopping condition for $\mu$=0.50 and 0.10.

### 3.2 Comparing the image reconstructions across the four methods

The SSIM and NRMSE values can be found in Table 1 for the raw data reconstructions (Supplemental Table S1 for the simulated data reconstructions) for all methods and acceleration factors. For display in subsequent figures, greater focus is placed on juxtaposing LR, CD, and CD+r as LR (NRMSE = 0.17 (±0.05), SSIM = 0.86 (±0.05)) performed better on average across subjects and $R$ values than L1 (NRMSE = 0.20 (±0.05), SSIM = 0.82 (±0.06)). Bolded values in Table 1 and Table S1 indicate the highest performing metrics, predominantly corresponding to CD and CD+r for $R \geq 18$.

Taking the same subject, S3, from Figure 2 with $\mu = 0.10$ for the CD+r reconstruction, Figure 3 compares the L1, LR, CD, and CD+r reconstructions for the simulated dataset (Figure 3A) and the raw dataset (Figure 3B) at $R$=12. In Figure 3A, L1 yields a higher NRMSE of 0.17 and a lower SSIM of 0.87, compared to NRMSE values of 0.13 for both LR, CD, and CD+r. The difference image in Figure 3A for L1 reveals a higher normalized error in the skull and grey matter regions, qualitatively visible as blurring in the magnified anatomical images. The results from the raw data analysis in Figure 3B are consistent with the results in Figure 3A. Averaged across all three raw datasets at $R$=12 in Figure 2B, the NRMSE values for the LR, CD, and CD+r reconstructions were 0.15 (±0.02), 0.14 (±0.02), and 0.14 (±0.02), respectively. Similarly, the average SSIM values for the three respective methods were 0.89 (±0.02), 0.88 (±0.03), and 0.88 (±0.03).

To compare the performance of LR, CD, and CD+r across datasets and acceleration factors, the SSIM values for $R \in \{8, 12, 18\}$ were plotted in Figure 4A against the flip angle (Supplemental Figure S7 for the simulations). The anatomical image of Subject S2 at $\theta$=10° is shown as a representative case in Figure 4B, where LR and CD reconstructions have larger SSIMs (+0.02) compared to CD+r for $R$=8 and $R$=12. At $R$=18, however, all four methods yield the same SSIM. In the SSIM plot for R=18, the CD and CD+r curves closely overlap and are higher than the L1 curve for flip angles below 12°. The anatomical images show that L1 reconstructions tend to be noisier, which obscures tissue details at higher acceleration factors compared to the smoother reconstructions from CD and CD+r. In the absence of model error, Supplemental Figure S7 shows that CD+r confers an advantage over CD for higher acceleration factors and stronger regularization.

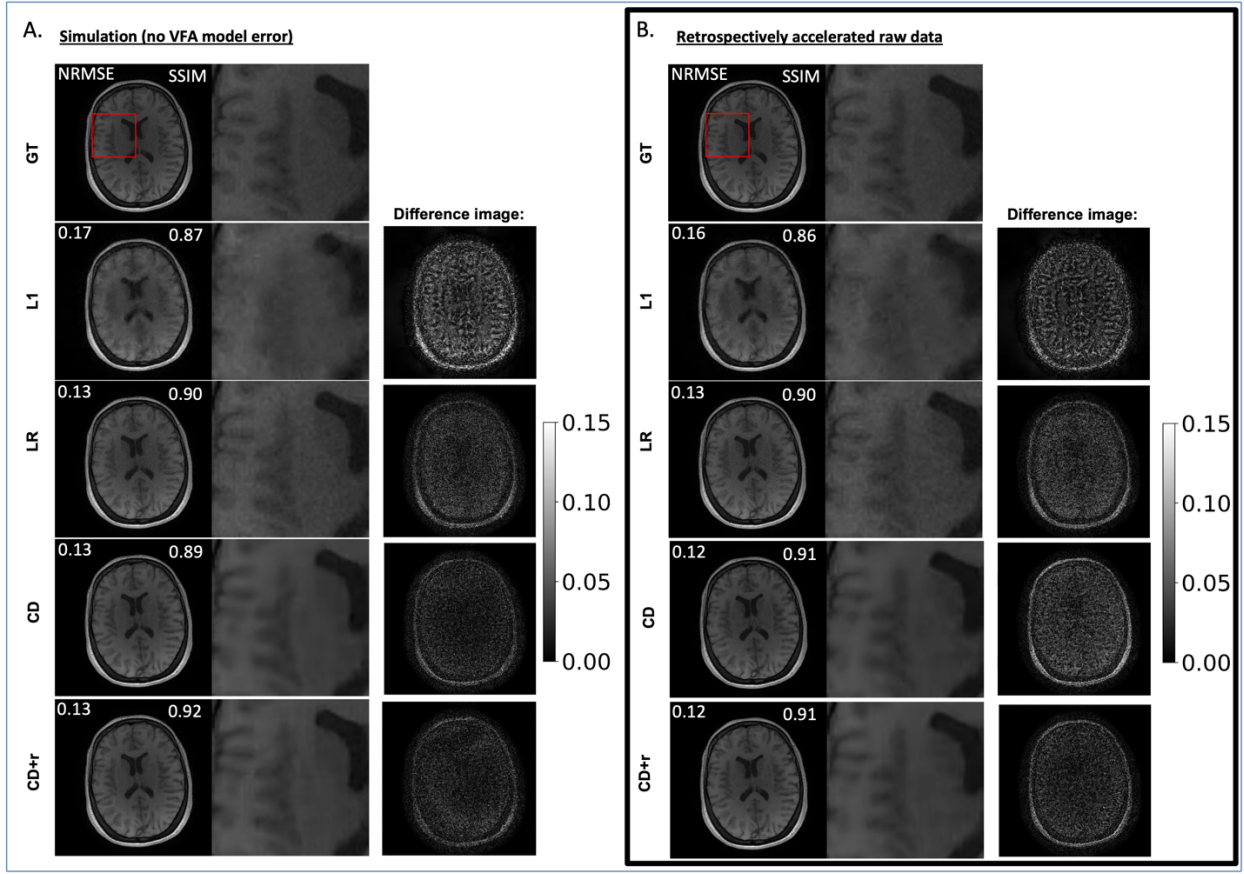

**Figure 3: Performance of all four methods in a representative dataset for $R = 12$.** A) Optimal reconstructions of simulated data at $\theta = 10°$ and $\mu = 0.10$ for CD+r. The CD+r reconstruction is compared to the results from L1, LR, and CD, where the white numbers overlayed on the images reflect the NRMSE and SSIM, as labeled, computed across all flip angles. The greyscale bar represents fractional error between the ground truth (GT) and each reconstruction. L1 yields a higher NRMSE of 0.17 and a lower SSIM of 0.87, compared to NRMSE values of 0.13 for the remaining methods. This higher error in the L1 reconstruction is evident in the magnified anatomical images, where there is loss of edge information in the grey matter region adjacent to the ventricle. In the absence of VFA model error, CD+r yielded the highest SSIM value among the reconstructions. (B) Analogous to the information in (A) for the raw data analysis. The results from the raw data analysis are consistent with the results in panel A, where all NRMSEs and SSIM values are within 0.1 of their simulation-based counterparts. CD+r performed similarly to CD when model error was no longer accounted for.

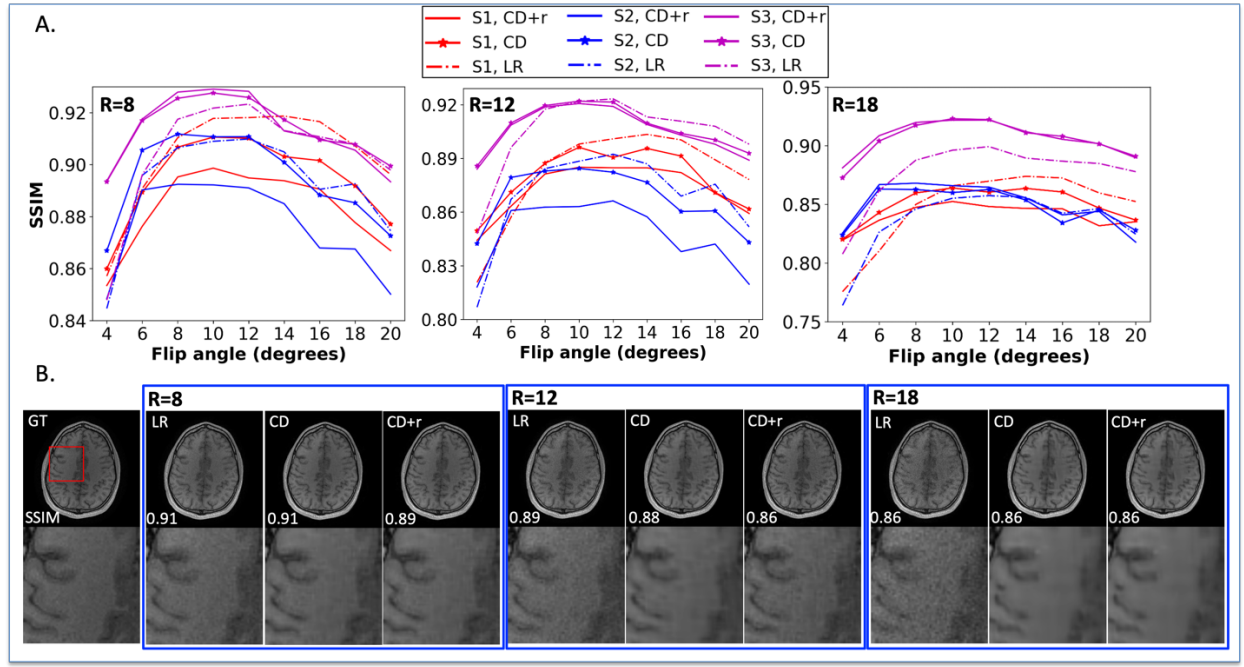

**Figure 4. ConvDecoder performance at different *R* values across all raw datasets**. (A) Plots of the SSIM as a function of flip angle for $R \in \{8, 12, 18\}$ for all three subjects. The solid, star, and dashed curves in the SSIM plots correspond to CD+r ($\mu = 0.10$), CD, and LR, respectively. (B) Anatomical images for subject S2 from each reconstruction (blue curves in panel (A)) at $\theta=10°$ for each of the three R values. The SSIM values are overlaid on each anatomical image. While LR and CD have higher SSIM values at the lowest two R values, the three methods match in performance at R=18, where all three blue curves in panel A closely overlap across all flip angles. The dynamic range in SSMI that the curves differ by with respect to one another does not exceed $\pm 0.2$ (within two significant figures).

*3.3 Comparing the $T_1$ maps across the four methods*

Aside from the reconstructed VFA image series from each of the four methods, we also investigated the quality of the $T_1$ maps corresponding to each optimal reconstruction. The NRMSEs and CCCs for the $T_1$ maps corresponding to the reconstructions from all accelerated raw data at all *R* can be found in Table 2 (simulation results in Supplemental Table S4). Bolded values in Table 2 and Supplemental Table S4 indicate the highest performing metrics, corresponding to CD and CD+r $T_1$ maps across all *R*.

Figure 5 compares the $T_1$ maps corresponding to the optimal LR, CD, and CD+r reconstructions for $R \in \{8, 12, 18\}$. Qualitatively, the LR $T_1$ map showed the least agreement with the ground truth $T_1$ map across all acceleration factors. For $R = 8, 12, 18$, respectively, the LR results were CCC = 0.91, 0.87, 0.83 compared to CD results of CCC = 0.93, 0.91, 0.88 and CD+r results of CCC = 0.93, 0.91, 0.89. At $R = 36$ (1.5 min of scan time), the CD+r $T_1$ map (CCC =

0.82) and the CD $T_1$ map (CCC=0.82) showcased similar agreement with the ground truth map. Across all acceleration factors, the percent difference between CCC values from the CD+r $T_1$ maps and the CD $T_1$ maps did not exceed 2%, while the largest percent difference encountered between the CD+r $T_1$ maps and the LR $T_1$ maps was 14% for $R = 36$.

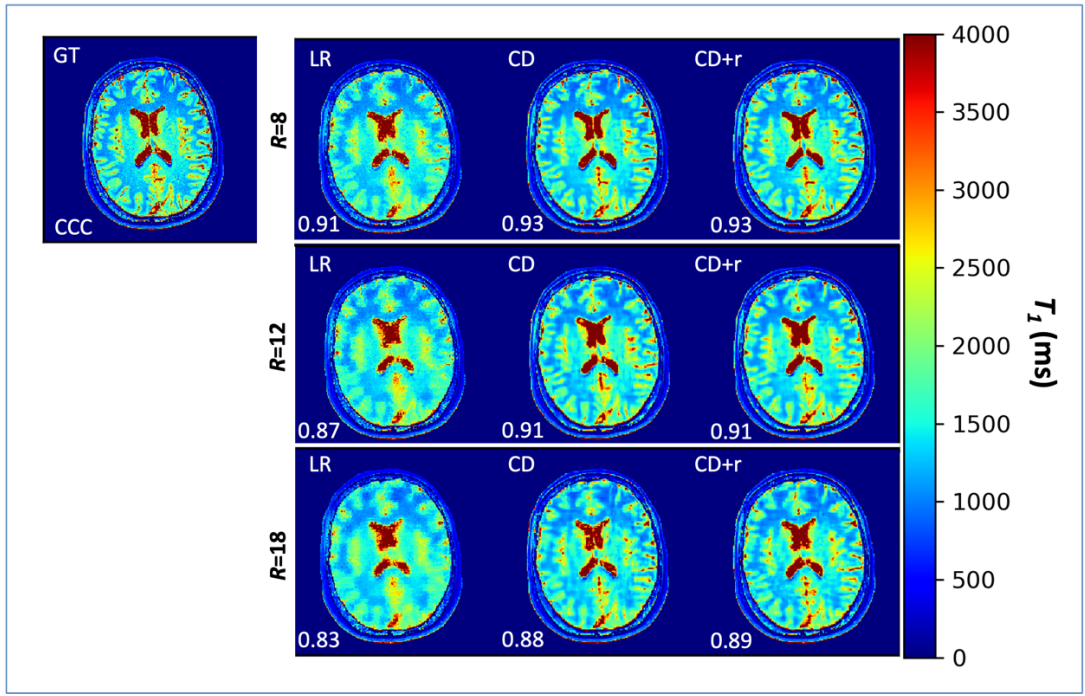

**Figure 5. $T_1$ maps for a single raw dataset at three accelerations**. $T_1$ maps for $R \in \{8, 12, 18\}$ corresponding to mapping the anatomical LR, CD, and CD+r reconstructions. The CD and CD+r ($\mu = 0.10$) reconstructions were obtained at the optimal number of training steps listed in Table 1. For all values of $R$, the LR T1 map has visible blurring in the grey matter folds compared to the CD and CD+r T1 maps. At $R=8$, the CCC value for the CD+r $T_1$ map is 2% higher than the CCC value for the L1 $T_1$ map; this difference is 7% at $R=18$ with CD+r showing greater agreement to the ground truth $T_1$ map.

Figure 6 displays the $T_1$ maps for all three retrospectively accelerated raw datasets at $R = 12$ (simulation results presented in Supplemental Figure S8). As in Figure 5, the LR $T_1$ maps lack spatially refinement across the tissue, most evident in the gray matter folds and ventricles that are clearly defined in the ground truth parameter map. Figure 7 compares the CCC and NRMSE values for the LR and CD+r $T_1$ maps from all subjects and acceleration factors, revealing that the CD+r $T_1$ maps across all experiments yield smaller NRMSE values ($p=0.162$) and larger CCC values ($p=0.131$) than the LR $T_1$ maps. The corresponding simulation results (Supplemental Figures S8 and S9) corroborated the raw data results, where the CD+r $T_1$ maps outperformed the LR $T_1$ maps to a statistically significant extent in terms of NRMSE (*p*=0.015) and CCC (*p*=0.002).

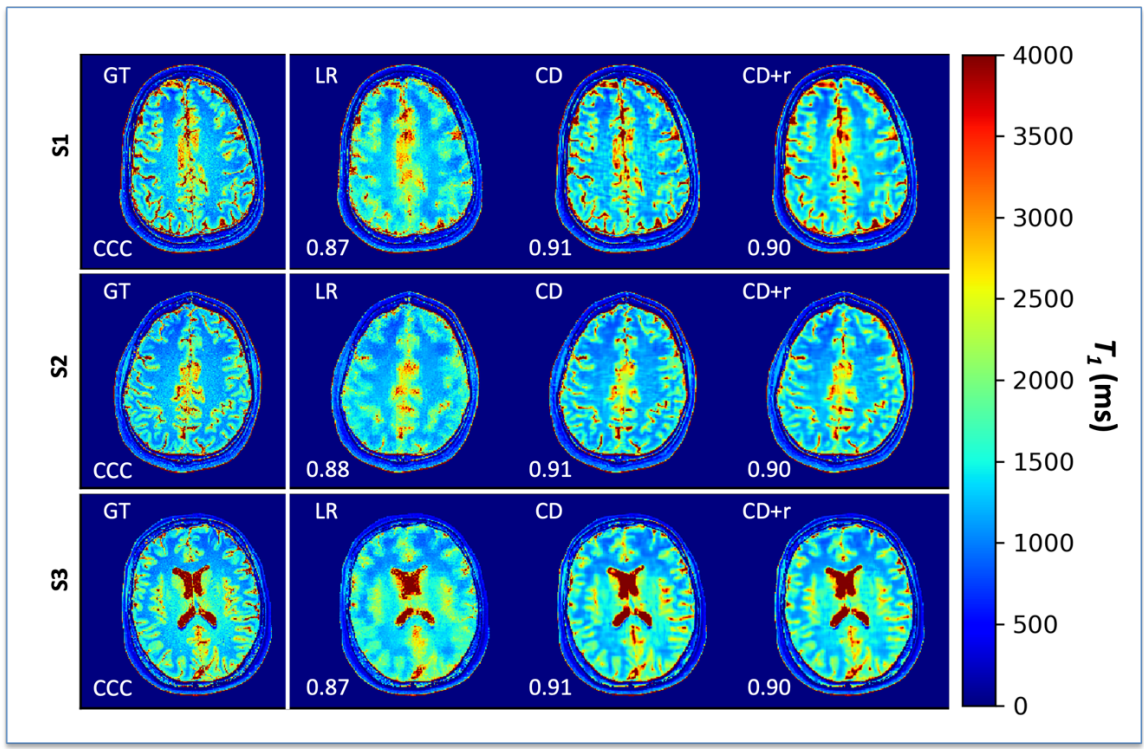

**Figure 6. $T_1$ maps for all raw datasets at $R$ = 12**. $T_1$ maps for all three subjects corresponding to mapping the anatomical LR, CD, and CD+r reconstructions. The CD and CD+r ($\mu$ = 0.10) reconstructions for each subject are displayed after the optimal number of training steps specified in Table 1. For all values of $R$, the CD and CD+r $T_1$ maps show less blurring in the grey matter regions as compared to the LR $T_1$ maps. The CCC values, overlain on each map, are lower for the LR $T_1$ maps for all subjects (≤0.88) as compared to the CD and CD+r $T_1$ maps for all subjects (≥ 0.90).

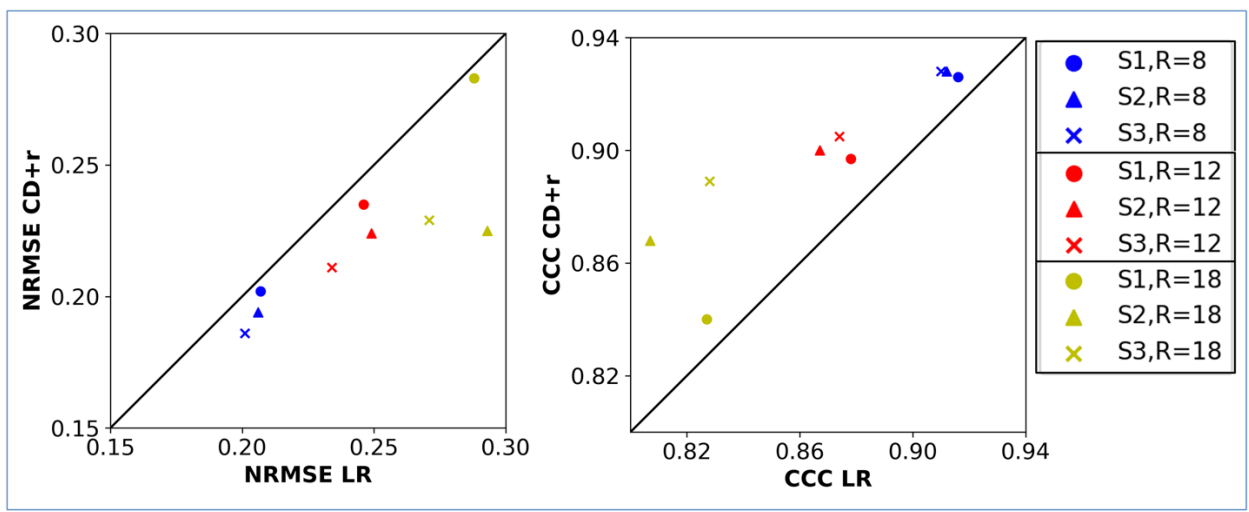

**Figure 7: NRMSE and CCC values for all $T_1$ maps from the raw data analysis**. The left plot compares the NRMSEs of the CD+r $T_1$ maps of all subjects against the NRMSEs of the corresponding LR $T_1$ maps across three *R* values. The right plot compares the CCCs of the CD+r $T_1$ maps of all subjects against the CCCs of the corresponding LR $T_1$ maps for the same three *R* values. The legend at left identifies the *R* values by color (blue: R=8, red: R=12, yellow: R=18) and the subjects by marker shape (circle: S1, triangle: S2, cross: S3). The LR NRMSE values are greater than the CD+r values (*p*=0.162) across all subjects and *R* values, and the LR CCC values are lower (*p*=0.131) in all cases.

## 4   DISCUSSION

This study is the first to utilize the untrained ConvDecoder method with a physics-based regularization term serving as a stopping condition to reconstruct retrospectively accelerated raw, dynamic MRI data, noting that the original CD architecture was previously only applied to reconstructing 2D static images with no physics-based regularization[31]. It is important to note that the NRMSE curve was known to us due to the retrospective nature of this study that provides access to ground truth data. We therefore used the NRMSE to choose the best stopping condition for L1, LR, and CD in order to provide a "best-case" reconstruction for comparison, noting that this is not achievable in practice. Importantly, CD+r was blind to the ground truth data, yet CD+r was able to perform similarly to the "best-case" result of CD, while outperforming the other two methods. While trained methods typically outperform untrained methods[31], it can be difficult to acquire ground truth training data for some types of scans, such as dynamic contrast-enhanced MRI and other quantitative imaging schemes that have strict temporal sampling requirements. There is therefore a need for untrained methods that can work with limited data.

A major challenge of using untrained methods is the selection of regularization parameter and early stopping condition that do not rely on having ground truth data to which to compare the reconstructed images at each training step. Importantly, the physics-based regularization proposed in this study demonstrated reliability as a stopping condition that can be computed without access to ground truth data given a suitable value for the regularization parameter. Even when neglecting early stopping and when trained to completion after 10,000 training steps, CD+r yielded reconstructions with substantially lower NRMSEs at the end of training compared to CD across all subjects, acceleration factors, and $\mu$ values. Because the optimal value of $\mu$ is not known *a priori* for novel reconstructions, we investigated $\mu$ values at different scales and subsequently showed that (i) the NRMSE increases slowly as training continues beyond the known (i.e. optimal) solution, and (ii) the automatic stopping condition remains close to the optimal solution. Thus, there is tolerance in the choice of solution based on the stopping condition and regularization parameter.

Another important contribution of the CD+r method is the ability to compute highly detailed quantitative parameter maps simultaneously during reconstruction *via* the physics-based regularization term. The $T_1$ maps corresponding to the CD+r reconstructions yielded high agreement with the reference $T_1$ map across all subjects, achieving CCC > 0.9 for R ≤ 12 (as short as 4-minute of acquisition time), thus outperforming the L1 and LR baselines. Comparing the CD+r $T_1$ maps with those derived from the baseline LR method (Figures 5 and 6), CD+r yields $T_1$ maps with more spatial refinement in closer agreement with the reference (GT) $T_1$ map. The LR $T_1$ map, however, exhibits greater noise and blurring, particularly in the gray matter region across subjects and acceleration factors. It is important to note that L1, LR, and CD benefited from *a priori* knowledge of the ground-truth when selecting the regularization parameters, while CD+r did not.

While CD+r compared positively to baseline methods, there are opportunities for further investigation. First, untrained methods are generally slower than trained methods, and CD+r is not excluded from this. Trained methods, while taking ample time and resources to train (particularly for large training datasets with thousands of examples), are fast at inference, outputting a solution after a small number of forward passes through a deep network, whereas, untrained methods are iteratively fit to data over many training steps for each independent set of data. As CD+r explicitly solves for the dynamic images, the reconstruction will require more time and memory for datasets with more dynamic points. One solution, proposed by Darestani *et al.*[31], is to warmstart the neural network (i.e., load pre-trained weights from a similar anatomy to initiate the network rather than beginning with random weights). Arora *et. al.* also showed that multiple slices could be reconstructed with a single generative model[58]. In Darestani *et al.*'s example, it was shown that a ConvDecoder architecture optimized for one set of non-fat suppressed knee MR images is a good initialization for training a ConvDecoder architecture to reconstruct another set of fat-suppressed knee MR images. In the context of this study, an adjacent tissue slice can be used to warmstart the CD+r reconstruction scheme for all subsequent brain reconstructions of the same imaging modality, and we show promising preliminary results for this approach in Supplemental Figure S11. It also may be possible to incorporate computation savings from subspace-based methods[59]. Additionally, because the physics-based regularization term in CD+r informs when the optimal solution is reached, training can be stopped early and would thus save up to one hour of computation (given the computational resources used in this work) compared to CD.

Another drawback to consider is that the current design of CD+r applies only to 2D dynamic data. As such, reconstructing a dynamic 3D volume would necessitate slice-by-slice reconstruction with warmstarting or an updated CD+r architecture that can take as input 4D, perhaps with convolutions along the dynamic dimension for further compression. This approach has successfully been explored in other trained methods for fast image reconstruction[60–62]. While we implemented one type of deep image prior, there remains the question of whether other DIPs, such as RARE[63] and TD-DIP[64], may confer a greater advantage. Similarly, the optimal sampling mask remains to be explored for our application. Lastly, as only healthy subjects were available for direct imaging in the research setting, there is the question of the performance of CD+r in the context of diseased tissue. This direction of work would involve close collaboration with radiologists to determine the diagnostic quality of images and close collaboration with mathematical oncologists to appropriately curate models of disease for regularization.

The analysis in this work indicated that it may be possible to use the unsupervised physics-based regularization term – blind to ground truth data – to determine the optimal stopping point in training while yielding results similar to methods that do reference the ground truth, which is an important real-world consideration when deploying these methods prospectively. These retrospective results indicate that it may be possible to achieve similar performance using the CD+r method given prospectively accelerated data, at least given the Poisson-disc sampling pattern used in this work. A prospective analysis merits further investigation.

While the CD+r method was evaluated using VFA data and the SPGR model, it is generalizable to any MRI modality and corresponding physical model, making it compatible with, for example, MRF, EPTI, and MR multitasking given the reformulation of the ConvDecoder network to output tensor data. In another example, a recent approach to choosing the stopping condition proposed splitting the measurements into three disjoint sets for training, validation, and early stopping, respectively; as this does not exploit the physical model, it could also be combined with the proposed physics-based regularization[65]. To recover high-spatial, high-temporal resolution images from an ultra-fast DCE-MRI acquisition[44], as another example, the dictionary used for fitting the SPGR model to the VFA data would be replaced with a dictionary storing perfusion model curves, such as those outputted by the Kety-Tofts[66] model. The structure and training of the network would be optimized for the DCE-MRI inverse problem through hyperparameter optimization of the number of layers, the size of the latent space, and the step size used for gradient descent.

## 5 CONCLUSION

We have introduced an untrained deep learning-based generative neural network for model-based MRI reconstruction with a physics-based regularization. We demonstrated the utility of the stopping condition provided by the physics-based regularization term in the CD+r training, which was especially powerful in $T_1$ mapping at high accelerations. The ability to determine the optimal reconstruction without a performance measure requiring ground truth data is integral to the reconstruction of prospectively acquired accelerated MRI data. The success of this method on highly accelerated raw data indicates the potential to incorporate a fast acquisition of dynamic MRI data in the clinical setting for quantitative imaging.

**ACKNOWLEDGEMENTS**

We thank the National Institutes of Health for funding through NCI U01CA142565, U01CA174706, U01CA253540, U24CA226110, and R01CA240589, as well as NIH U24EB029240. We also extend our gratitude to the American Cancer Society for support through RSG-18-006-01-CCE and the Cancer Prevention and Research Institute of Texas for support through CPRIT RR160005. An Amazon Web Services Machine Learning Research Award was an integral funding component in enabling the use of GPUs. Lastly, we gratefully acknowledge the time that the three subjects in this study volunteered for being scanned. T.E.Y. is a CPRIT Scholar in Cancer Research.

**REFERENCES:**

1. Kim DS, Kong MH, Jang SY, Kim JH, Kang DS, Song KY. The Usefulness of Brain Magnetic Resonance Imaging with Mild Head Injury and the Negative Findings of Brain Computed Tomography. *J Korean Neurosurg Soc*. 2013;54(2):100. doi:10.3340/jkns.2013.54.2.100
2. Smajlović D, Sinanović O. Sensitivity of the neuroimaging techniques in ischemic stroke. *Med Arh*. 2004;58(5):282-284.
3. Kertesz A, Black SE, Nicholson L, Carr T. The sensitivity and specificity of MRI in stroke. *Neurology*. 1987;37(10):1580-1585. doi:10.1212/wnl.37.10.1580

4. Kuhl C, Weigel S, Schrading S, et al. Prospective multicenter cohort study to refine management recommendations for women at elevated familial risk of breast cancer: The EVA trial. *Journal of Clinical Oncology*. 2010;28(9):1450-1457. doi:10.1200/JCO.2009.23.0839
5. Kuhl CK, Schrading S, Strobel K, Schild HH, Hilgers RD, Bieling HB. Abbreviated breast Magnetic Resonance Imaging (MRI): First postcontrast subtracted images and maximum-intensity projection - A novel approach to breast cancer screening with MRI. *Journal of Clinical Oncology*. 2014;32(22):2304-2310. doi:10.1200/JCO.2013.52.5386
6. Leithner D, Moy L, Morris EA, Marino MA, Helbich TH, Pinker K. Abbreviated MRI of the Breast: Does It Provide Value? *Journal of Magnetic Resonance Imaging*. 2019;49(7):e85-e100. doi:10.1002/jmri.26291
7. Shehata M, Grimm L, Ballantyne N, et al. Ductal Carcinoma in Situ: Current Concepts in Biology, Imaging, and Treatment. *J Breast Imaging*. 2019;1(3):166-176. doi:10.1093/jbi/wbz039
8. Abramson RG, Arlinghaus LR, Dula AN, et al. MR Imaging Biomarkers in Oncology Clinical Trials. *Magn Reson Imaging Clin N Am*. 2016;24(1):11-29. doi:10.1016/j.mric.2015.08.002
9. Rosenkrantz AB, Mendiratta-Lala M, Bartholmai BJ, et al. Clinical Utility of Quantitative Imaging. *Acad Radiol*. 2015;22(1):33-49. doi:10.1016/j.acra.2014.08.011
10. Yankeelov TE, Mankoff DA, Schwartz LH, et al. Quantitative imaging in cancer clinical trials. *Clinical Cancer Research*. 2016;22(2):284-290. doi:10.1158/1078-0432.CCR-14-3336
11. Wang HZ, Riederer SJ, Lee JN. Optimizing the precision in T1 relaxation estimation using limited flip angles. *Magn Reson Med*. 1987;5(5):399-416. doi:10.1002/mrm.1910050502
12. Müller A, Jurcoane A, Kebir S, et al. Quantitative T1-mapping detects cloudy-enhancing tumor compartments predicting outcome of patients with glioblastoma. *Cancer Med*. 2017;6(1):89-99. doi:10.1002/cam4.966
13. Yankeelov T, Gore J. Dynamic Contrast Enhanced Magnetic Resonance Imaging in Oncology:Theory, Data Acquisition,Analysis, and Examples. *Curr Med Imaging Rev*. 2007;3(2):91-107. doi:10.2174/157340507780619179
14. Calcagno C, Lobatto ME, Dyvorne H, et al. Three-dimensional dynamic contrast-enhanced MRI for the accurate, extensive quantification of microvascular permeability in atherosclerotic plaques. *NMR Biomed*. 2015;28(10):1304-1314. doi:10.1002/nbm.3369
15. Bergamino M, Barletta L, Castellan L, Mancardi G, Roccatagliata L. Dynamic Contrast-Enhanced MRI in the Study of Brain Tumors. Comparison Between the Extended Tofts-Kety Model and a Phenomenological Universalities (PUN) Algorithm. *J Digit Imaging*. 2015;28(6):748-754. doi:10.1007/s10278-015-9788-2
16. Sorace AG, Barnes SL, Hippe DS, et al. Distinguishing benign and malignant breast tumors: preliminary comparison of kinetic modeling approaches using multi-institutional dynamic contrast-enhanced MRI data from the International Breast MR Consortium 6883 trial. *Journal of Medical Imaging*. 2018;5(01):1. doi:10.1117/1.jmi.5.1.011019
17. Virostko J, Hainline A, Kang H, et al. Dynamic contrast-enhanced magnetic resonance imaging and diffusion-weighted magnetic resonance imaging for predicting the response of locally advanced breast cancer to neoadjuvant therapy: a meta-analysis. *Journal of Medical Imaging*. 2017;5(01):1. doi:10.1117/1.JMI.5.1.011011

18. Jarrett AM, Kazerouni AS, Wu C, et al. Quantitative magnetic resonance imaging and tumor forecasting of breast cancer patients in the community setting. *Nat Protoc*. 2021;16(11):5309-5338. doi:10.1038/s41596-021-00617-y
19. Virostko J, Sorace AG, Slavkova KP, et al. Quantitative multiparametric MRI predicts response to neoadjuvant therapy in the community setting. *Breast Cancer Research*. 2021;23(1):110. doi:10.1186/s13058-021-01489-6
20. Ma D, Gulani V, Seiberlich N, et al. Magnetic resonance fingerprinting. *Nature*. 2013;495(7440):187-192. doi:10.1038/nature11971
21. Poorman ME, Martin MN, Ma D, et al. Magnetic resonance fingerprinting Part 1: Potential uses, current challenges, and recommendations. *J Magn Reson Imaging*. 2020;51(3):675-692. doi:10.1002/jmri.26836
22. Christodoulou AG, Shaw JL, Nguyen C, et al. Magnetic resonance multitasking for motion-resolved quantitative cardiovascular imaging. *Nat Biomed Eng*. 2018;2(4):215-226. doi:10.1038/s41551-018-0217-y
23. Ma S, Nguyen CT, Han F, et al. Three-dimensional simultaneous brain $T_1$, $T_2$, and ADC mapping with MR Multitasking. *Magn Reson Med*. 2020;84(1):72-88. doi:10.1002/mrm.28092
24. Wang N, Xie Y, Fan Z, et al. Five-dimensional quantitative low-dose Multitasking dynamic contrast- enhanced MRI: Preliminary study on breast cancer. *Magn Reson Med*. 2021;85(6):3096-3111. doi:10.1002/mrm.28633
25. Wang F, Dong Z, Reese TG, et al. Echo planar time-resolved imaging (EPTI). *Magn Reson Med*. 2019;81(6):3599-3615. doi:10.1002/mrm.27673
26. Dong Z, Wang F, Reese TG, Bilgic B, Setsompop K. Echo planar time-resolved imaging with subspace reconstruction and optimized spatiotemporal encoding. *Magn Reson Med*. 2020;84(5):2442-2455. doi:10.1002/mrm.28295
27. Fessler JA. MODEL-BASED IMAGE RECONSTRUCTION FOR MRI. *IEEE Signal Process Mag*. 2010;27(4):81-89. doi:10.1109/MSP.2010.936726
28. Aggarwal HK, Mani MP, Jacob M. MoDL: Model-Based Deep Learning Architecture for Inverse Problems. *IEEE Trans Med Imaging*. 2019;38(2):394-405. doi:10.1109/TMI.2018.2865356
29. Hammernik K, Klatzer T, Kobler E, et al. Learning a variational network for reconstruction of accelerated MRI data. *Magn Reson Med*. 2018;79(6):3055-3071. doi:10.1002/mrm.26977
30. Yaman B, Hosseini SAH, Moeller S, Ellermann J, Ugurbil K, Akcakaya M. Ground-truth free multi-mask self-supervised physics-guided deep learning in highly accelerated MRI. In: *Proceedings - International Symposium on Biomedical Imaging*. Vol 2021-April. IEEE Computer Society; 2021:1850-1854. doi:10.1109/ISBI48211.2021.9433924
31. Zalbagi Darestani M, Heckel R. Accelerated MRI With Un-Trained Neural Networks. *IEEE Trans Comput Imaging*. 2021;7:724-733. doi:10.1109/TCI.2021.3097596
32. Heckel R, Hand P. Deep Decoder: Concise Image Representations from Untrained Non-convolutional Networks. *arXiv preprint*. Published online October 2, 2018.
33. van Veen D, Jalal A, Soltanolkotabi M, Price E, Vishwanath S, Dimakis AG. Compressed Sensing with Deep Image Prior and Learned Regularization. *arXiv preprint*. Published online June 17, 2018.
34. Bora A, Jalal A, Price E, Dimakis AG. Compressed Sensing using Generative Models. *arXiv preprint*. Published online March 9, 2017.

35. Ulyanov D, Vedaldi A, Lempitsky V. Deep Image Prior. *Int J Comput Vis*. 2020;128(7):1867-1888. doi:10.1007/s11263-020-01303-4
36. Liu F, Kijowski R, el Fakhri G, Feng L. Magnetic resonance parameter mapping using model-guided self-supervised deep learning. *Magn Reson Med*. 2021;85(6):3211-3226. doi:10.1002/mrm.28659
37. Ronneberger O, Fischer P, Brox T. U-Net: Convolutional Networks for Biomedical Image Segmentation. In: Navab N, Hornegger J, Wells WM, Frangi AF, eds. *Lecture Notes in Computer Science: Medical Image Computing and Computer-Assisted Intervention – MICCAI 2015*. Vol 9351. Springer, Cham; 2015:234-241. doi:10.1007/978-3-319-24574-4_28
38. Yaman B, Amir S, Hosseini H, Akçakaya M. Zero-Shot Physics-Guided Deep Learning for Subject-Specific MRI Reconstruction. In: *NeurIPS Workshop on Deep Learning and Inverse Problems*. ; 2021.
39. Leynes AP, Nagarajan SS, Larson PEZ. Scan-specific Self-supervised Bayesian Deep Non-linear Inversion for Undersampled MRI Reconstruction. *ArXiv*. Published online March 1, 2022.
40. Slavkova KP, DiCarlo JC, Wadhwa V, et al. Implementing ConvDecoder with physics-based regularization to reconstruct under-sampled variable-flip angle MRI data of the breast. In: *Proceedings of the Annual Meeting of the ISMRM*. International Society for Magnetic Resonance in Medicine; 2021:1448. Accessed February 5, 2022. https://index.mirasmart.com/ISMRM2021/PDFfiles/1448.html
41. Zhang T, Pauly JM, Levesque IR. Accelerating parameter mapping with a locally low rank constraint. *Magn Reson Med*. 2015;73(2):655-661. doi:10.1002/mrm.25161
42. Lustig M, Donoho D, Pauly JM. Sparse MRI: The application of compressed sensing for rapid MR imaging. *Magn Reson Med*. 2007;58(6):1182-1195. doi:10.1002/mrm.21391
43. Pruessmann KP, Weiger M, Scheidegger MB, Boesiger P. SENSE: sensitivity encoding for fast MRI. *Magn Reson Med*. 1999;42(5):952-962.
44. Buxton RB, Edelman RR, Rosen BR, Wismer GL, Brady TJ. Contrast in Rapid MR Imaging. *J Comput Assist Tomogr*. 1987;11(1):7-16. doi:10.1097/00004728-198701000-00003
45. Tamir JI, Yu SX, Lustig M. DeepInPy: Deep Inverse Problems in Python. In: *ISMRM Workshop on Data Sampling and Image Reconstruction*. ; 2020.
46. Keenan KE, Ainslie M, Barker AJ, et al. Quantitative magnetic resonance imaging phantoms: A review and the need for a system phantom. *Magn Reson Med*. 2018;79(1):48-61. doi:10.1002/mrm.26982
47. Kingma DP, Ba J. Adam: A Method for Stochastic Optimization. *ArXiv*. Published online December 22, 2014.
48. Savitzky Abraham, Golay MJE. Smoothing and Differentiation of Data by Simplified Least Squares Procedures. *Anal Chem*. 1964;36(8):1627-1639. doi:10.1021/ac60214a047
49. Guo S, Fessler JA, Noll DC. High-Resolution Oscillating Steady-State fMRI Using Patch-Tensor Low-Rank Reconstruction. *IEEE Trans Med Imaging*. 2020;39(12):4357-4368. doi:10.1109/TMI.2020.3017450
50. BART Toolbox for Computational Magnetic Resonance Imaging. doi:DOI:10.5281/zenodo.592960
51. Buehrer M, Pruessmann KP, Boesiger P, Kozerke S. Array compression for MRI with large coil arrays. *Magn Reson Med*. 2007;57(6):1131-1139. doi:10.1002/mrm.21237

52. Zhang T, Pauly JM, Vasanawala SS, Lustig M. Coil compression for accelerated imaging with Cartesian sampling. *Magn Reson Med*. 2013;69(2):571-582. doi:10.1002/mrm.24267
53. Uecker M, Lai P, Murphy MJ, et al. ESPIRiT-an eigenvalue approach to autocalibrating parallel MRI: Where SENSE meets GRAPPA. *Magn Reson Med*. 2014;71(3):990-1001. doi:10.1002/mrm.24751
54. Lustig M, Donoho D, Santos J. Compressed sensing MRI. *Signal Processing Magazine, IEEE*. 2008;(March 2008):72-82. doi:Doi 10.1109/Tit.2006.871582
55. Trzasko J, Manduca A. CLEAR : Calibration-Free Parallel Imaging using Locally Low-Rank Encouraging ReconstructionCLEAR : Calibration-Free Parallel Imaging using Locally Low-Rank Encouraging Reconstruction. In: *Proceedings of the 20th Annual Meeting of ISMRM*. ; 2012:517.
56. Lin LIK. A Concordance Correlation Coefficient to Evaluate Reproducibility. *Biometrics*. 1989;45(1):255. doi:10.2307/2532051
57. Wang Z, Bovik AC, Sheikh HR, Simoncelli EP. Image Quality Assessment: From Error Visibility to Structural Similarity. *IEEE Transactions on Image Processing*. 2004;13(4):600-612. doi:10.1109/TIP.2003.819861
58. Arora S, Roeloffs V, Lustig M. Untrained Modified Deep Decoder for Joint Denoising and Parallel Imaging Reconstruction. In: *Proceedings of the International Society for Magnetic Resonance in Medicine*. ; 2020.
59. Tamir JI, Uecker M, Chen W, et al. $T_2$ shuffling: Sharp, multicontrast, volumetric fast spin-echo imaging. *Magn Reson Med*. 2017;77(1):180-195. doi:10.1002/mrm.26102
60. Qin C, Schlemper J, Caballero J, Price AN, Hajnal J v., Rueckert D. Convolutional Recurrent Neural Networks for Dynamic MR Image Reconstruction. *IEEE Trans Med Imaging*. 2019;38(1):280-290. doi:10.1109/TMI.2018.2863670
61. Küstner T, Fuin N, Hammernik K, et al. CINENet: deep learning-based 3D cardiac CINE MRI reconstruction with multi-coil complex-valued 4D spatio-temporal convolutions. *Sci Rep*. 2020;10(1):13710. doi:10.1038/s41598-020-70551-8
62. Sandino CM, Lai P, Vasanawala SS, Cheng JY. Accelerating cardiac cine MRI using a deep learning-based ESPIRiT reconstruction. *Magn Reson Med*. 2021;85(1):152-167. doi:10.1002/mrm.28420
63. Liu J, Sun Y, Eldeniz C, Gan W, An H, Kamilov US. RARE: Image Reconstruction using Deep Priors Learned without Ground Truth. Published online December 12, 2019. doi:10.1109/JSTSP.2020.2998402
64. Yoo J, Jin KH, Gupta H, Yerly J, Stuber M, Unser M. Time-Dependent Deep Image Prior for Dynamic MRI. *IEEE Trans Med Imaging*. 2021;40(12):3337-3348. doi:10.1109/TMI.2021.3084288
65. Yaman B, Amir Hossein Hosseini S, Akcakaya M. Zero-Shot Self-Supervised Learning for MRI Reconstruction . In: *International Conference on Learning Representation*. OpenReview; 2022.
66. Tofts PS, Kermode AG. Measurement of the blood-brain barrier permeability and leakage space using dynamic MR imaging. 1. Fundamental concepts. *Magn Reson Med*. 1991;17(2):357-367. doi:10.1002/mrm.1910170208

| R=8 | | | | | | | | |
|---|---|---|---|---|---|---|---|---|
| Subject | L1 | | LR | | CD | | CD+r (μ = 0.10) | |
| | Performance | Regularization | Performance | Regularization | Performance | Number of steps | Performance | Number of steps |
| S1 | (0.158, 0.862) | 0.01 | (0.141, **0.904**) | 0.0079 | (**0.139**,0.894) | 1702 (1500) | (0.149,0.883) | 1414 (1500) |
| S2 | (0.155, 0.864) | 0.01 | (0.137, **0.892**) | 0.0077 | (**0.134**,0.876) | 1695 (1500) | (0.136, 0.876) | 1934 (2000) |
| S3 | (0.135, 0.885) | 0.01 | (0.116, **0.921**) | 0.0073 | (**0.111**, 0.914) | 2060 (2000) | (**0.111**, 0.913) | 2196 (2000) |
| **R=12** | | | | | | | | |
| Subject | L1 | | LR | | CD | | CD+r (μ = 0.10) | |
| | Performance | Regularization | Performance | Regularization | Performance | Number of steps | Performance | Number of steps |
| S1 | (0.179, 0.844) | 0.01 | (0.161, **0.882**) | 0.0063 | (**0.149**,0.879) | 1695 (1500) | (0.163,0.873) | 1391 (1500) |
| S2 | (0.177, 0.841) | 0.0098 | (0.155, **0.869**) | 0.006 | (**0.143**,0.868) | 1645 (1500) | (0.149,0.848) | 1485 (1500) |
| S3 | (0.157, 0.864) | 0.0091 | (0. 131, 0.904) | 0.0058 | (**0.119**, **0.907**) | 2156 (2000) | (0.120, 0.906) | 1947 (2000) |
| **R=18** | | | | | | | | |
| Subject | L1 | | LR | | CD | | CD+r (μ = 0.10) | |
| | Performance | Regularization | Performance | Regularization | Performance | Number of steps | Performance | Number of steps |
| S1 | (0.210, 0.811) | 0.0078 | (0.189, 0.848) | 0.0049 | (**0.162**,0.880) | 1798 (2000) | (0.198,0.840) | 773 (1000) |
| S2 | (0.207, 0.802) | 0.0071 | (0.182, 0.835) | 0.0046 | (**0.151**,0.848) | 1728 (1500) | (0.163,**0.850**) | 1440 (1500) |
| S3 | (0.186, 0.836) | 0.0065 | (0.155, 0.877) | 0.0045 | (**0.13**,0.907) | 2262 (2500) | (**0.130**, 0.907) | 2967 (3000) |
| **R=36** | | | | | | | | |
| Subject | L1 | | LR | | CD | | CD+r (μ = 0.10) | |
| | Performance | Regularization | Performance | Regularization | Performance | Number of steps | Performance | Number of steps |
| S1 | (0.289, 0.726) | 0.0038 | (0.256, 0.775) | 0.0048 | (**0.199**, **0.837**) | 2177 (2000) | (0.339, 0.681) | 9772 (9999) |
| S2 | (0.277, 0.718) | 0.0036 | (0.249, 0.752) | 0.0051 | (**0.185**, **0.817**) | 2030 (2000) | (0.307, 0.701) | 7660 (7500) |
| S3 | (0.261, 0.756) | 0.003 | (0.219, 0.810) | 0.0038 | (**0.155**, **0.864**) | 2486 (2500) | (0.156, 0.866) | 2551 (2500) |

Table 1. Performance measures, (NRMSE, SSIM), for all reconstruction methods applied to raw data. The optimal number of iterations for L1 and LR were 110 and 150, respectively.

| R=8 | | | | |
|---|---|---|---|---|
| Subject | L1 | LR | CD | CD+r ($\mu = 0.10$) |
| S1 | (0.204, 0.921) | (0.207, 0.916) | **(0.201, 0.928)** | (0.202, 0.926) |
| S2 | (0.201, 0.918) | (0.206, 0.912) | (0.204, 0.920) | **(0.194, 0.928)** |
| S3 | (0.193, 0.918) | (0.201, 0.910) | **(0.185, 0.928)** | (0.186, **0.928**) |
| R=12 | | | | |
| Subject | L1 | LR | CD | CD+r ($\mu = 0.10$) |
| S1 | (0.243, 0.883) | (0.246, 0.878) | **(0.219, 0.914)** | (0.235, 0.897) |
| S2 | (0.248, 0.869) | (0.249, 0.867) | **(0.220, 0.907)** | (0.224, 0.900) |
| S3 | (0.230, 0.879) | (0.234, 0.874) | **(0.211, 0.906)** | (**0.211**, 0.905) |
| R=18 | | | | |
| Subject | L1 | LR | CD | CD+r ($\mu = 0.10$) |
| S1 | (0.0.287, 0.830) | (0.0.288, 0.827) | **(0.254, 0.883)** | (0.283, 0.840) |
| S2 | (0.290, 0.813) | (0.293, 0.807) | **(0.249, 0.876)** | (0.255, 0.868) |
| S3 | (0.265, 0.837) | (0.271, 0.828) | (0.235, 0.880) | **(0.229, 0.889)** |
| R=36 | | | | |
| Subject | L1 | LR | CD | CD+r ($\mu = 0.10$) |
| S1 | (0.364, 0.703) | (0.374, 0.682) | **(0.315, 0.811)** | (0.434, 0.633) |
| S2 | (0.38, 0.624) | (0.408, 0.575) | **(0.301, 0.813)** | (0.417, 0.648) |
| S3 | (0.329, 0.739) | (0.332, 0.730) | (0.280, **0.822**) | (**0.277**, 0.821) |

Table 2. Performance measures for $T_1$ mapping analysis of the raw data. Paired values in parentheses in each field correspond to the (NRMSE, CCC).

*Supplemental Materials*

**An untrained deep learning method for reconstructing dynamic magnetic resonance images from accelerated model-based data**

Kalina P. Slavkova, Julie C. DiCarlo, Viraj Wadhwa, Sidharth Kumar, Chengyue Wu, John Virostko, Thomas E. Yankeelov, Jonathan I. Tamir

Supplemental Methods

*Data processing*

After raw k-space datasets were collected as described in the main text, the data were processed and structured into *h5py* files using Python to conform with data formatting standards in DeepInPy. A representative slice from each dataset was selected by qualitatively examining the magnitude images. Subjects S1 and S2 exhibited RF artifacts in most central slices, leading to the selection of more superior slices. S3, however, exhibited no visible artifacts and enabled the selection of the central slice containing the ventricles and cerebrospinal fluid. All data and images are complexed values, and this was handled in Python and PyTorch by defining a real and complex channel. Thus, the *k*-space of each subject was stored with dimensions [batch, coils, flip angles, X, Y, complex/real channels] (where batch = 1 as each dataset is optimized over independently), and images were stored with dimensions [batch, flip angles, X, Y, complex/real channels]. The figure below displays an example complex-valued image computed from the fully sampled k-space of a representative subject, where the flip angles ascend from left to right (4°, 6°, 8°, …20°).

Real:

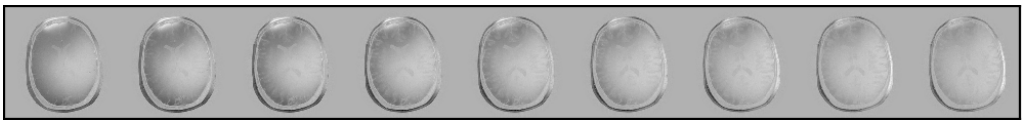

Imaginary:

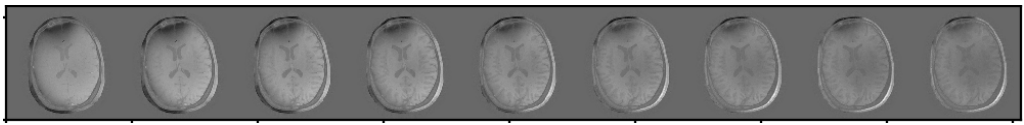

Magnitude:

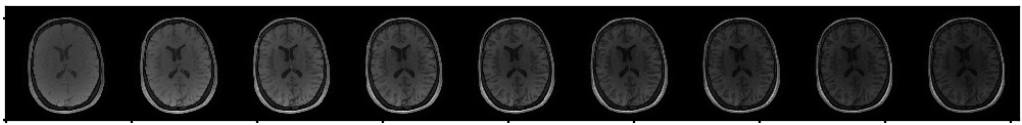

Phase:

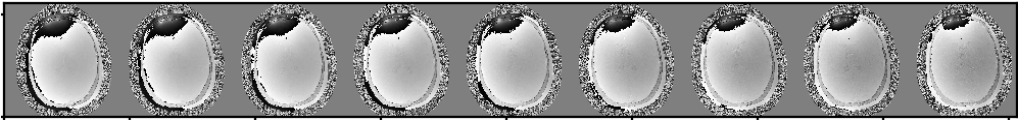

The *h5py* files consisted of four fields: 'ksp' (storing the fully sampled k-space data), 'imgs' (storing the images computed from the k-space), 'maps' (storing the sensitivity maps computed as described in the main text using BART), and 'masks' (storing the Poisson-disc sampling mask with a specified acceleration factor). The DeepInPy software loads the h5py file, applies the mask to the data to arrive at under-sampled k-space, and defines a forward operator A using the masks and maps. While the original ConvDecoder paper reconstructed coil-images, the modified network applied in this study reconstructs complex coil-combined images.

*Phantom scans*

As described in the main texts, the phantom was scanned with a protocol similar to that used for acquiring the subject datasets. The following figure shows the phantom images at each flip angle in the VFA series, where the flip angles ascend from left to right (4°, 6°, 8°, …20°).

Real:

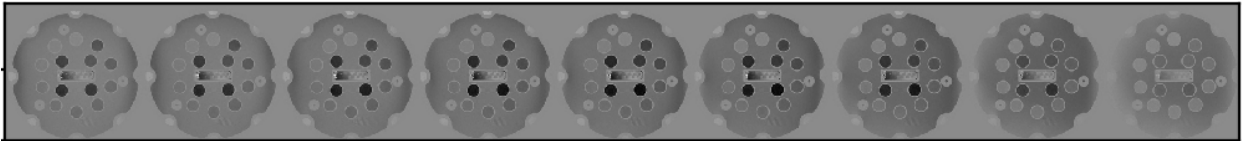

Imaginary:

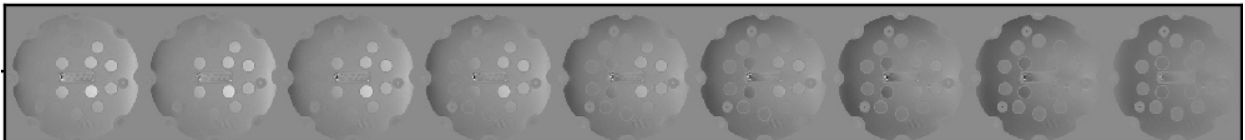

Magnitude:

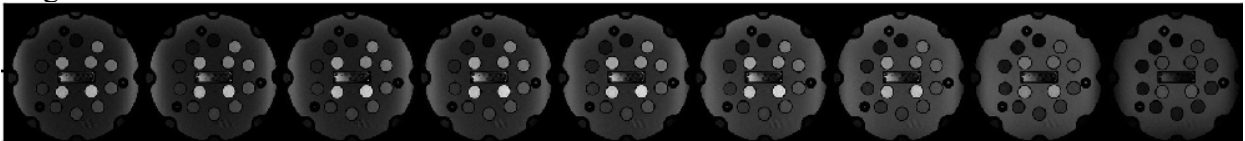

Phase:

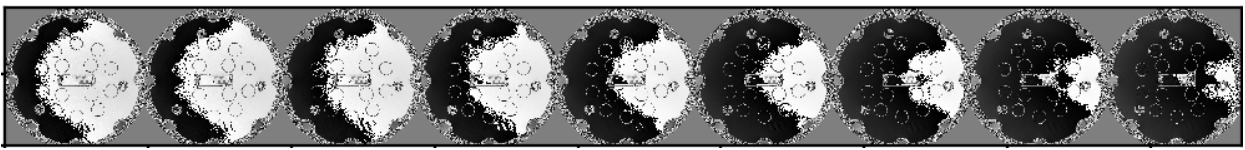

All hyperparameter optimization was performed on phantom data. The following table details the hyperparameters and the corresponding range of values that were considered in the optimization. As described in the main text, the phantom data experiments were also used to determine an appropriate norm for the data, which we empirically found to be 1,000. This was deduced by qualitatively inspecting the loss curves for smoothness for a few different norms (10, 100, 1000).

| | |
|---|---|
| Number of blocks | [4, 6, 8] |
| Latent channels | [64, 96, 128, 256] |
| Input dimension | [4, 6, 8, 64] |
| Step size | [0.1, 0.01, 0.001, 0.0001] |
| Data norm | [10, 100, 1000] |

*Dictionary matching*

To fit VFA imaging data computed from the acquired data to the signal model, we implemented a dictionary matching technique that is commonly employed in the quantitative MRI literature (as in MRF). As described in the main text, we built our dictionary by running the signal model forward for a vector of 2,000 T1 values linearly spaced between 50 to 2000 ms. We investigated the effect of the number of simulated curves, *N*, in the dictionary on the accuracy of the fit, and we found that a dictionary built from as few as 50 linearly spaced T1 values is accurate (see figure below). The NRMSE of the fitted VFA images from a dictionary with N=50 was 0.0680, compared to an NRMSE of 0.0676 for *N*=2,000 (used for the results of this manuscript). Thus, we are confident in the accuracy of our dictionary matching technique and found no need to increase the number of linearly spaced T1 values from 2,000.

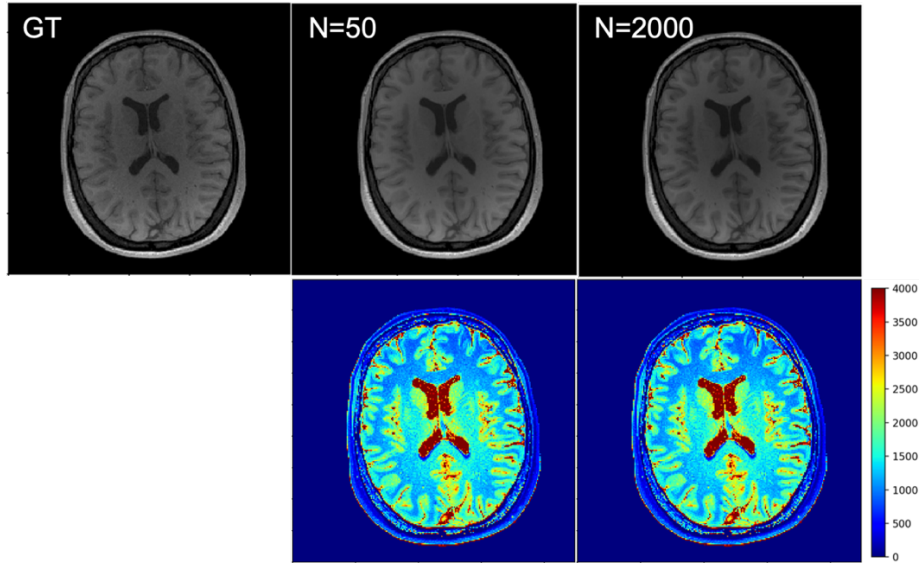

The dictionary matching scheme implemented in our methodology is based on minimizing the following equation:

$$x_m = \max_{T1} \|x^T D(T1, \theta)\|$$

In the context of CD+r optimization, $x=G(w)$, where $x_m$ is updated with $x=G(w)$ every five optimization steps as described in Section 2.3 of the Methods in the main body. In the Python implementation of our dictionary matching algorithm, the max is computed over the index of the T1 vector, and the T1 map is arrived at by indexing the T1 vector with these indices. The T1 map is not used in the optimization and is, rather, a quantitative output acquired in addition to the reconstructed images.

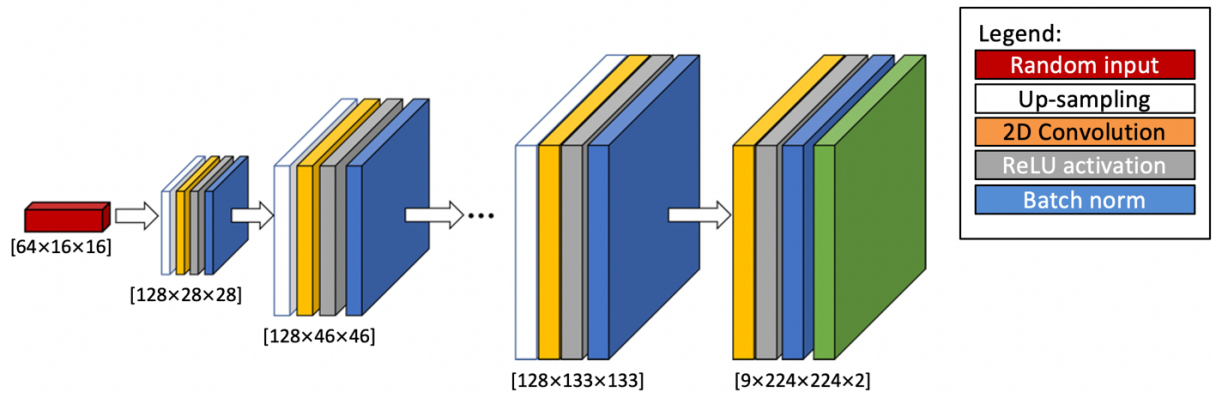

**Figure S1: Modified ConvDecoder architecture for multi-contrast MRI reconstruction**. The ConvDecoder architecture consists of six blocks with 128 latent channels. From left to right (denoted by the white arrows) the input of random noise (red tetrahedron of size 64×16×16, Gaussian noise with mean 0 and standard deviation of 1) is operated on by the first layer of the architecture. The output of each layer serves as the input into the next layer until the final complex-valued VFA images (green) are outputted from the final layer of size 9×224×224×2. Each layer (excluding the last) consists of a sequential order of operations as follows: up-sampling operation (white) to increase the size of the input, a convolutional operation (orange) consisting of kernels of size 3×3, a nonlinearity conferred by ReLU activation functions (dark gray), and batch normalization (dark blue). The final layer does not up-sample its input and instead directly convolves the input, passes it through the nonlinearity, performs batch normalization, and outputs an image with the user-specified dimensions. The size of each layer output is as follows: [128×28×28], [128×46×46], [128×78×78], [128×133×133], and [9×224×224×2].

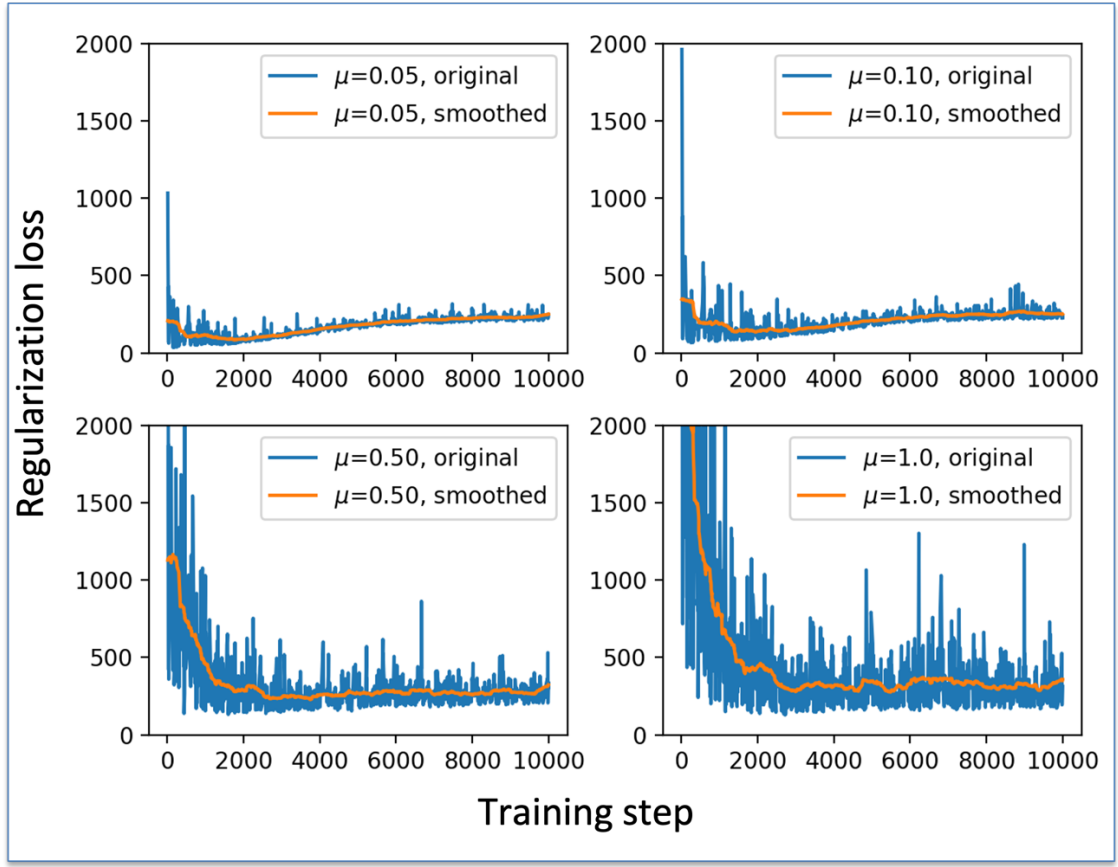

**Figure S2: Regularization loss for raw dataset S3 at R = 12 for all regularization parameters.** Each plot displays the actual regularization loss (blue) plotted across all 10,000 training steps with the smoothed regularization loss (orange) overlaid after application of the Savitzky-Golay method, which is effectively a low-pass filter. This smoothing method functions by filtering the signal with an *n*-degree polynomial of a fixed window size, where the polynomial is fit at each point in the signal, using least-squares with derivate of order *m*, capturing surrounding points within the extent of the window. For smoothing the regularization loss and NRMSE in this work, a window size of 51 was chosen with $n = 1$ and $m = 0$.

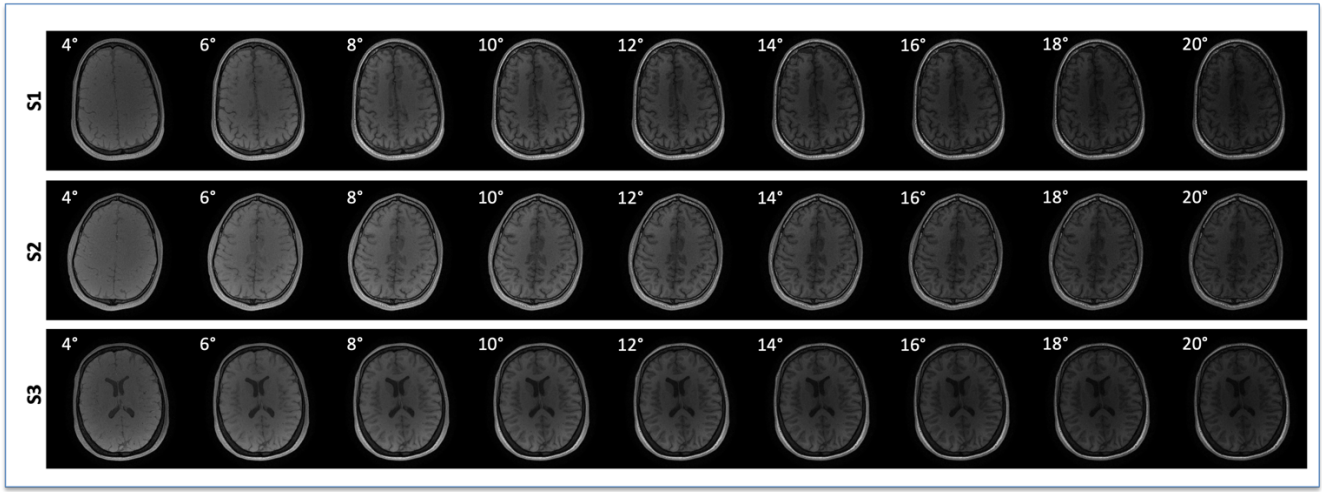

**Figure S3: Ground truth anatomical images from fully-sampled raw data**. The $T_1$-weighted, anatomical images at all nine flip angles are displayed for each subject (S1, S2, S3) for reference. The signal intensity is lower for higher flip angles, and the dynamic range of the contrast is greater for mid-range flip angles.

Submitted to *Magnetic Resonance in Medicine* (response to reviewers submitted)

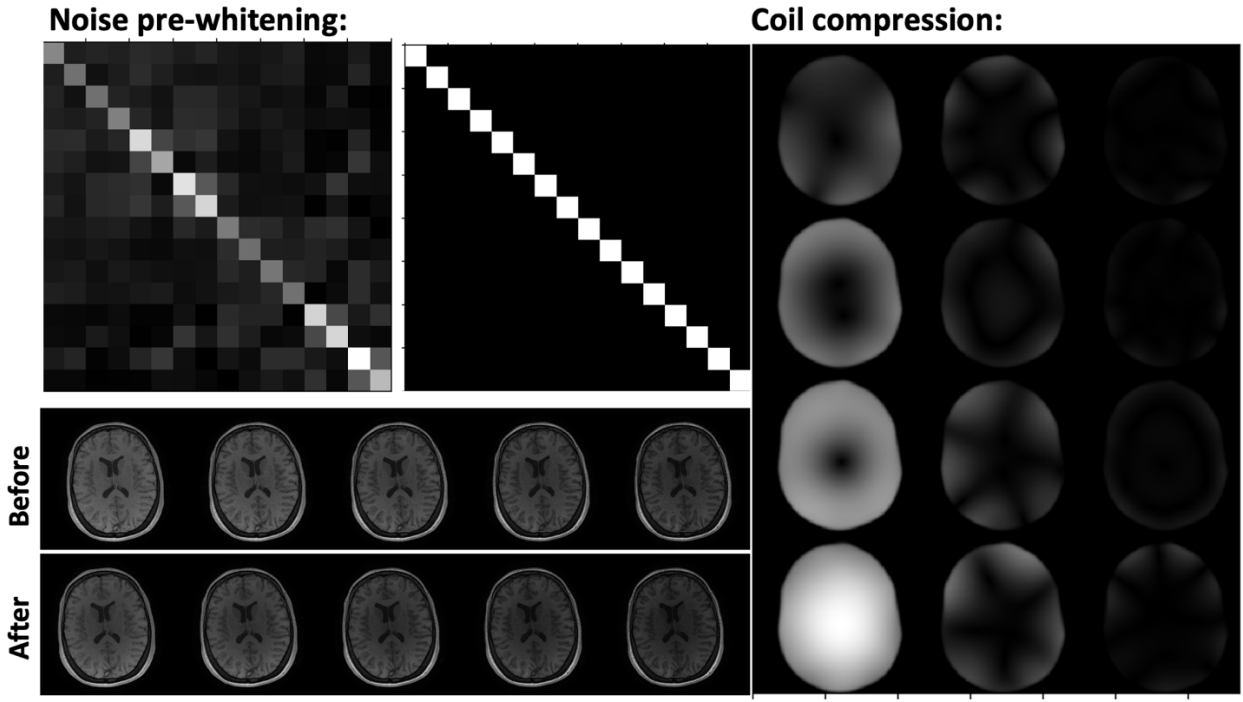

**Figure S4: Data pre-processing steps**. The correlation matrix between all 16 original sensitivity coils is shown in the top left panel, demonstrating that there is correlation (non-zero off-diagonal elements of the matrix) between the coils. After noise pre-whitening is applied to each *k*-space dataset, we see that the off-diagonal correlations have been eliminated (before and after images are shown for completeness). Following noise pre-whitening, software coil compression based on singular value decomposition was applied to solve for 11 signal-contributing virtual coils.

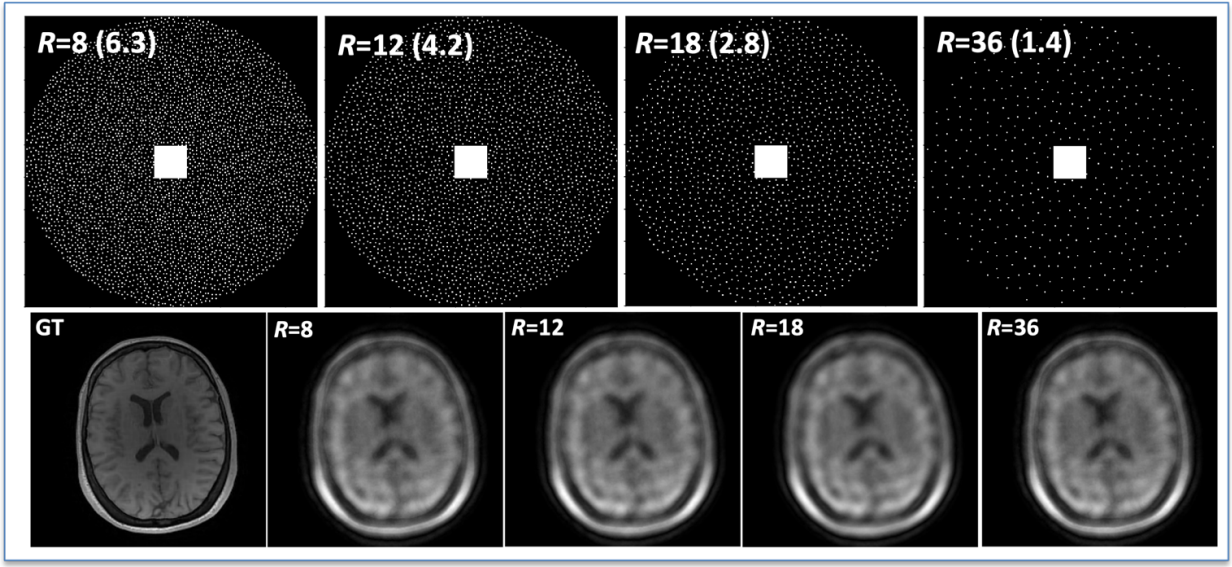

**Figure S5: Sampling masks and corresponding accelerated zero-filled images**. From the top left to the top right, the Poisson disk sampling masks are displayed for each of the four acceleration factors, $R \in \{8, 12, 18, 36\}$, with the corresponding equivalent scan time in minutes in parentheses. The anatomical images spanning from the bottom left to the bottom right of the figure are the corresponding ground-truth (GT) and zero-filled images for a representative scan.

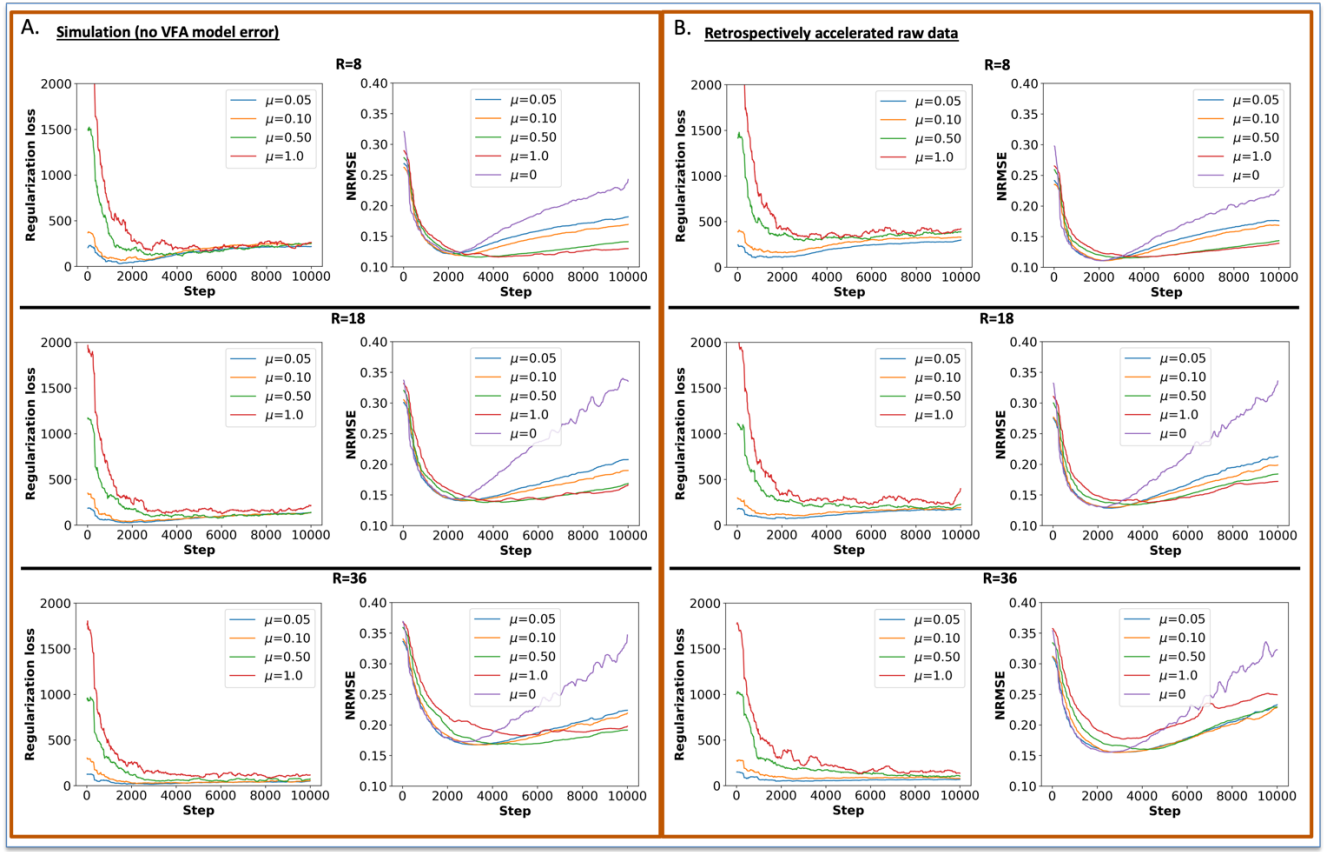

**Figure S6: Regularization loss and NRMSE in a representative dataset for $R \in \{8, 18, 36\}$.** A) The left and right plots in each row, separated by a black line, correspond to a particular $R$ value; these plots show the regularization loss (left) and the NRMSE (right) as a function of training step for the simulated data. (B) Analogous to plots in (A) for the raw data. The regularization loss curves corresponding to the two smallest regularization rates of 0.05 and 0.10 are more amendable to qualitative inspection of the minima, especially for $R = 36$.

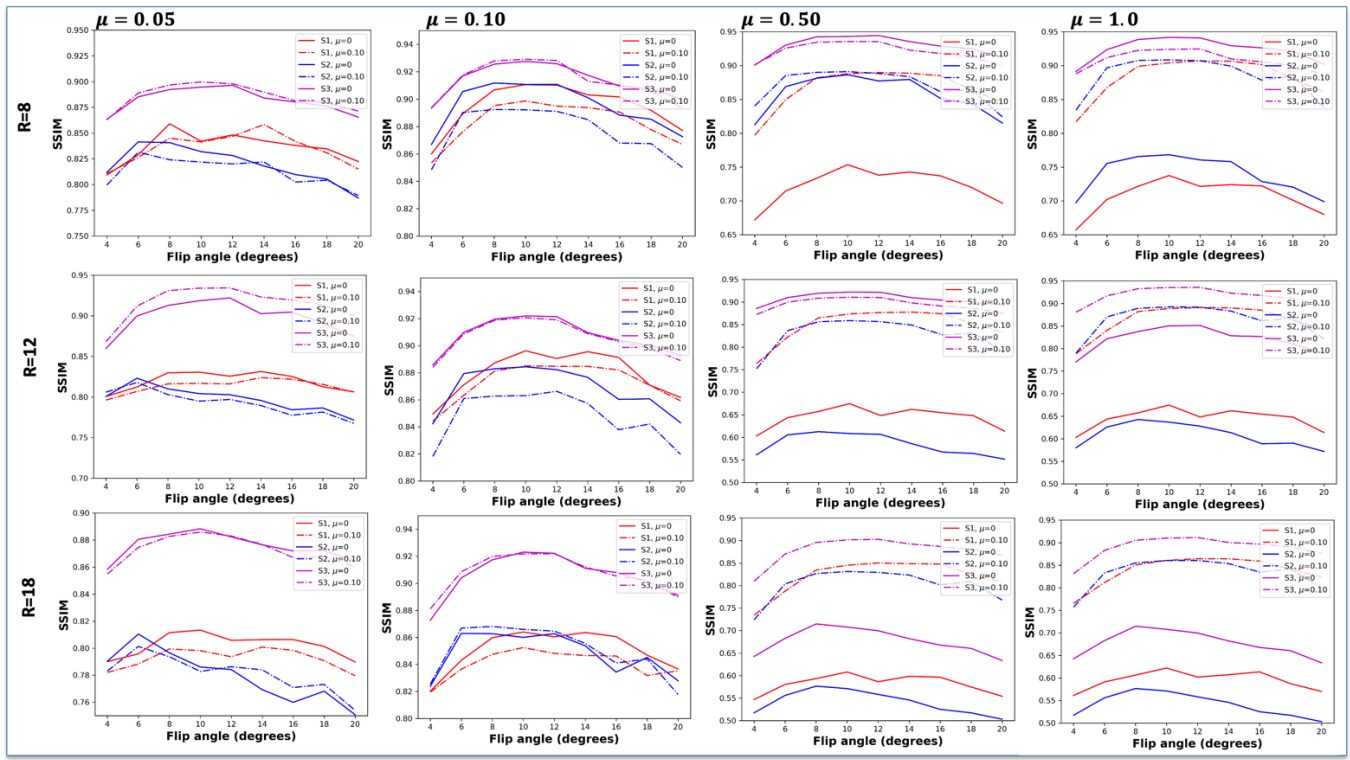

**Figure S7. CD and CD+r performance at different R values across all simulated datasets**. (A) Plots of the SSIM as a function of flip angle for $R \in \{8, 12, 18\}$ for all three subjects for four regularization rates in CD+r. The solid and dashed curves in the SSIM plots correspond to CD and CD+r, respectively.

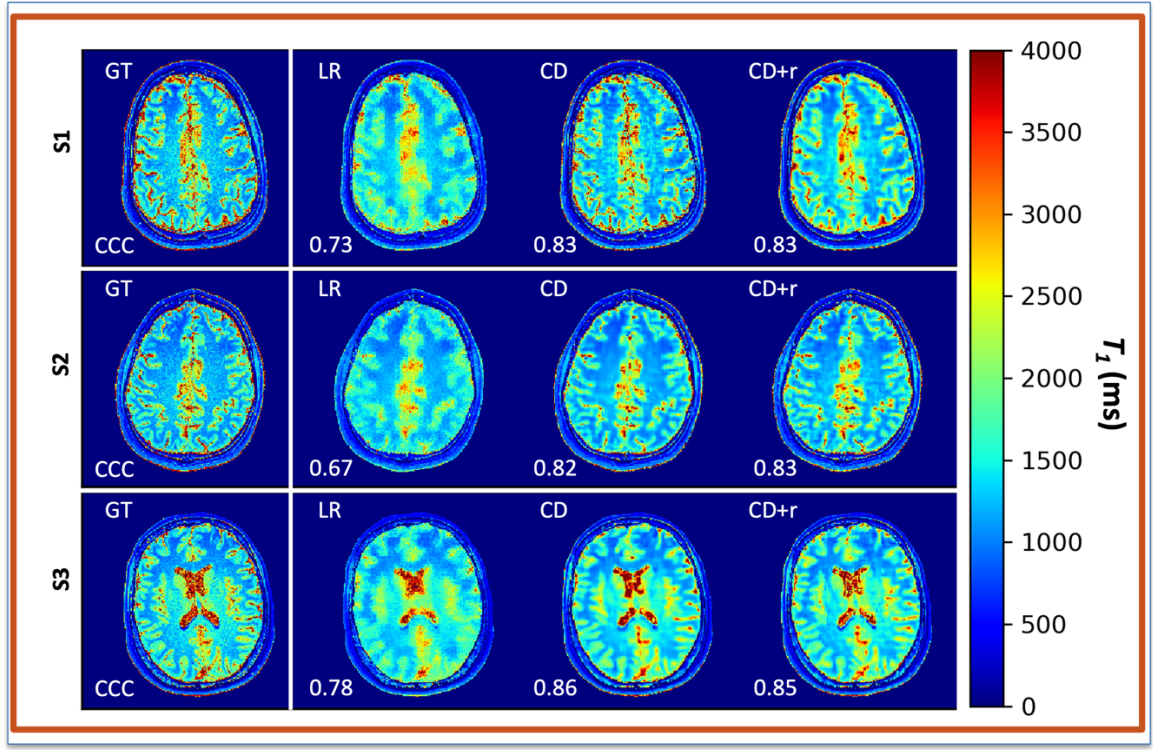

**Figure S8. *T₁* maps for all simulated datasets at R = 12**. $T_1$ maps for all three subjects corresponding to mapping the anatomical LR, CD, and CD+r reconstructions. The CD and CD+r ($\mu$=0.10) reconstructions for each subject were saved at their optimal number of training steps specified in Table 1. For all R's, CD and CD+r recover more spatially refined $T_1$ values in the gray matter and ventricles than LR. The CCC values, overlain on each map, are lower for the LR $T_1$ maps for all subjects (CCC<0.78) as compared to the CD and CD+r $T_1$ maps for all subjects (CCC>0.82).

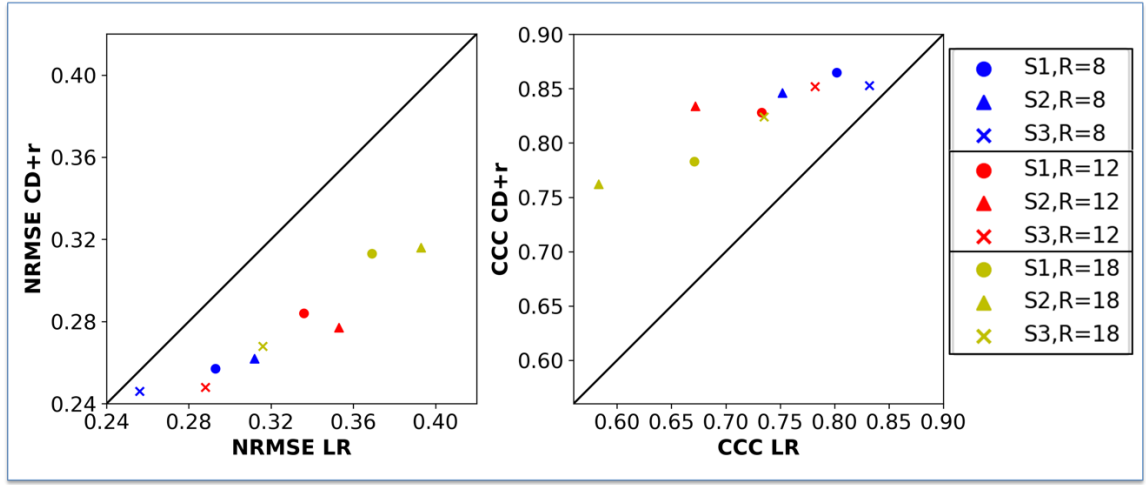

**Figure S9. NRMSE and CCC values for all $T_1$ maps from the simulated data analysis**. The left plot compares the NRMSEs across R∈{8, 12, 18} of the CD+r $T_1$ maps of each subject against the NRMSEs of the corresponding LR $T_1$ maps. The right plot compares the CCCs (for the same R values) of the CD+r $T_1$ maps of each subject against the CCCs of the three corresponding LR $T_1$ maps. The legend at left identifies the R values by color (blue: R=8, red: R=12, yellow: R=19) and the subjects by marker shape (circle: S1, triangle: S2, cross: S3). For all subjects and acceleration factors, the CD+r $T_1$ maps have lower NRMSE (*p*=0.015) and higher CCC (*p*=0.002) than the LR $T_1$ maps.

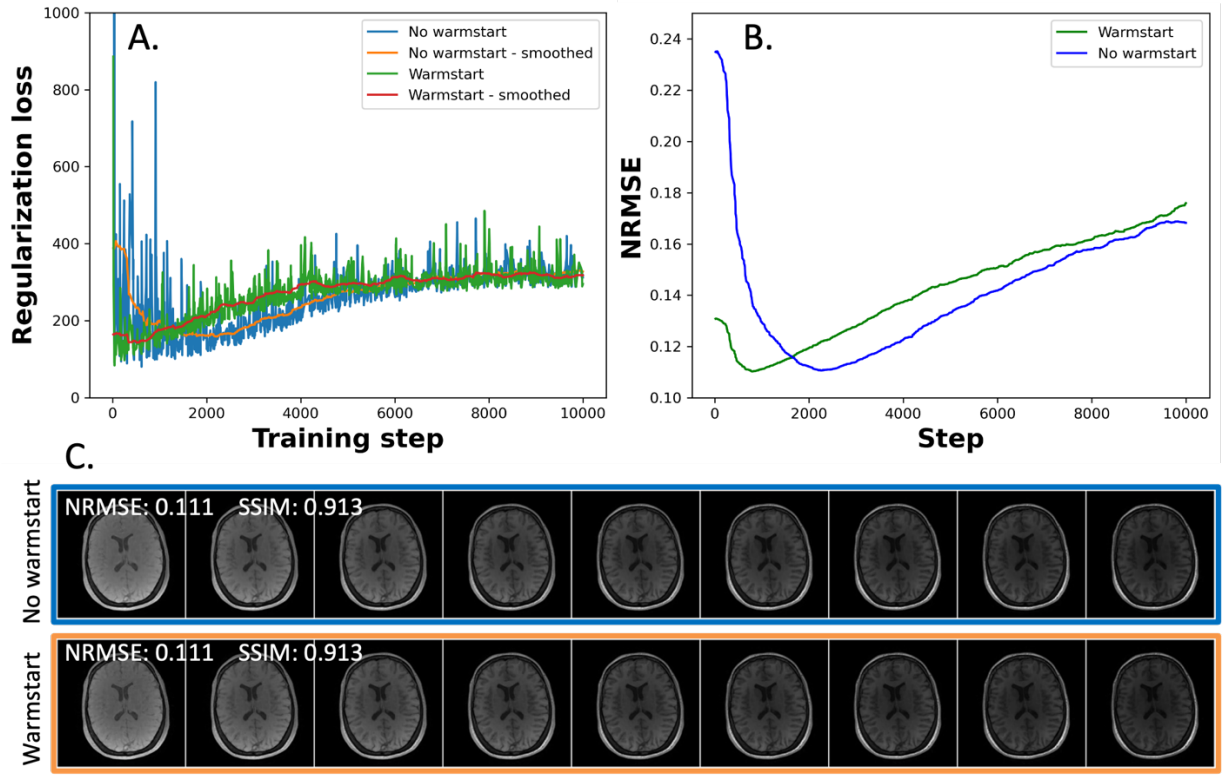

**Figure S11. Warm starting training using a neighboring slice in a representative dataset.** (A) Regularization loss for the original reconstruction with randomly initialized weights (blue and orange curves, where the orange curve is a smoothed version of the blue curve) and the warmstarted reconstruction (green and red curves, where the red curve is a smoothed version of the green curve). (B) NRMSE for the randomly initialized (blue) and warmstarted network (green). (C) Optimal reconstructions from the case of no warmstarting (blue-boxed image) at 1,947 optimization steps and with warm starting (orange-boxed figure) at 601 optimization steps. The NRMSE and SSIM of both reconstructions were the same. These results demonstrate the ability to save computation time (324% in this case) by initializing CD+r with the optimal weights saved form reconstructing a neighboring slice in the tissue volume.

Supplemental Tables

| R=8 | Simulation | | | | | | | |
|---|---|---|---|---|---|---|---|---|
| Subject | L1 | | LR | | CD | | CD+r (µ=0.10) | |
|  | Performance | Regularization | Performance | Regularization | Performance | Optimal steps | Performance | Optimal steps |
| S1 | (0.176, 0.865) | 0.01 | (0.141, 0.900) | 0.0086 | (0.142, **0.909**) | 1853 (2000) | (**0.140**, 0.887) | 1745 (1500) |
| S2 | (0.178, 0.861) | 0.01 | (0.144, 0.885) | 0.0084 | (**0.141, 0.858**) | 1645 (1500) | (0.141, 0.854) | 1699 (1500) |
| S3 | (0.150, 0.886) | 0.01 | (**0.119**, 0.918) | 0.0076 | (0.124, **0.928**) | 2547 (2500) | (0.129, 0.899) | 1489 (1500) |
| R=12 | | | | | | | | |
| Subject | L1 | | LR | | CD | | CD+r (µ=0.10) | |
|  | Performance | Regularization | Performance | Regularization | Performance | Optimal steps | Performance | Optimal steps |
| S1 | (0.199, 0.846) | 0.0094 | (0.155, 0.881) | 0.0066 | (0.157, **0.902**) | 1775 (2000) | (**0.154**, 0.875) | 1737 (1500) |
| S2 | (0.203, 0.837) | 0.0089 | (0.158, **0.864**) | 0.0065 | (0.156, 0.853) | 1637 (1500) | (**0.153, 0.864**) | 1789 (2000) |
| S3 | (0.173, 0.865) | 0.0084 | (0.132, 0.901) | 0.0058 | (0.132, 0.892) | 2085 (2000) | (**0.130, 0.916**) | 2600 (2500) |
| R=18 | | | | | | | | |
| Subject | L1 | | LR | | CD | | CD+r (µ=0.10) | |
|  | Performance | Regularization | Performance | Regularization | Performance | Optimal steps | Performance | Optimal steps |
| S1 | (0.230, 0.819) | 0.0069 | (0.176, 0.852) | 0.0049 | (**0.169, 0.870**) | 1702 (1500) | (0.172, 0.861) | 1675 (1500) |
| S2 | (0.234, 0.804) | 0.0065 | (0.179, 0.832) | 0.0048 | (0.171, **0.867**) | 1830 (2000) | (**0.165**, 0,827) | 2060 (2000) |
| S3 | (0.202, 0.839) | 0.0061 | (0.153, 0.874) | 0.0044 | (**0.143, 0.892**) | 2498 (2500) | (0.146, 0.878) | 1947 (2000) |
| R=36 | | | | | | | | |
| Subject | L1 | | LR | | CD | | CD+r (µ=0.10) | |
|  | Performance | Regularization | Performance | Regularization | Performance | Optimal steps | Performance | Optimal steps |
| S1 | (0.309, 0.750) | 0.0033 | (0.247, 0.770) | 0.0035 | (0.216, **0.846**) | 2254 (2500) | (**0.212**, 0.834) | 2881 (3000) |
| S2 | (0.306, 0.734) | 0.0031 | (0.248, 0.750) | 0.0035 | (0.207, 0.809) | 2237 (2000) | (**0.201, 0.826**) | 2881 (3000) |
| S3 | (0.278, 0.771) | 0.0027 | (0.221, 0.800) | 0.0033 | (**0.172, 0.849**) | 2710 (3000) | (0.177, 0.835) | 2239 (2000) |

Table S1. Performance measures, (NRMSE, SSIM), for all reconstruction methods applied to the simulated data.

| R=8 | Raw data | | | | | | | |
|---|---|---|---|---|---|---|---|---|
| Subject | CD+r (μ=0.05) | | CD+r (μ=0.10) | | CD+r (μ=0.50) | | CD+r (μ=1.0) | |
| | Performance | Optimal steps | Performance | Optimal steps | Performance | Optimal steps | Performance | Optimal steps |
| S1 | (0.220,0.835) | 421 (500) | (**0.149, 0.883**) | 1414 (1500) | (0.207, 0.866) | 9621 (9500) | (0.202, 0.883) | 9766 (9999) |
| S2 | (0.207, 0.813) | 421 (500) | (**0.136, 0.876**) | 1934 (2000) | (0.147,0.870) | 2881 (3000) | (0.182,0.884) | 7155 (7000) |
| S3 | (0.131, 0.885) | 808 (1000) | (**0.111**, 0.913) | 2196 (2000) | (0.115,**0.921**) | 2967 (3000) | (0.117,0.919) | 3772 (4000) |
| R=12 | | | | | | | | |
| Subject | CD+r (μ=0.05) | | CD+r (μ=0.10) | | CD+r (μ=0.50) | | CD+r (μ=1.0) | |
| | Performance | Optimal steps | Performance | Optimal steps | Performance | Optimal steps | Performance | Optimal steps |
| S1 | (0.206,0.813) | 573 (500) | (**0.163,0.873**) | 1391 (1500) | (0.222,0.848) | 8843 (9000) | (0.214,0.863) | 9084 (9000) |
| S2 | (0.175,0.793) | 671 (500) | (**0.149,0.848**) | 1485 (1500) | (0.228,0.828) | 9996 (9999) | (0.212,0.863) | 8340 (8500) |
| S3 | (0.121, 0.904) | 1789 (2000) | (**0.120**, 0.906) | 1947 (2000) | (0.125, **0.915**) | 3759 (4000) | (0.130, 0.916) | 5495 (5500) |
| R=18 | | | | | | | | |
| Subject | CD+r (μ=0.05) | | CD+r (μ=0.10) | | CD+r (μ=0.50) | | CD+r (μ=1.0) | |
| | Performance | Optimal steps | Performance | Optimal steps | Performance | Optimal steps | Performance | Optimal steps |
| S1 | (0.198,0.840) | 773 (1000) | (**0.179,0.840**) | 1244 (1000) | (0.246,0.820) | 8290 (8500) | (0.231,0.838) | 7346 (7500) |
| S2 | (0.198,0.781) | 649 (500) | (**0.163,0.850**) | 1440 (1500) | (0.249,0.802) | 9353 (9500) | (0.244,0.833) | 9547 (9500) |
| S3 | (0.135, 0.873 ) | 1675 (1500) | (**0.130, 0.907**) | 2967 (3000) | (0.182,0.878) | 9621 (9500) | (0.170, 0.889) | 9538 (9500) |
| R=36 | | | | | | | | |
| Subject | CD+r (μ=0.05) | | CD+r (μ=0.10) | | CD+r (μ=0.50) | | CD+r (μ=1.0) | |
| | Performance | Optimal steps | Performance | Optimal steps | Performance | Optimal steps | Performance | Optimal steps |
| S1 | (0.342, 0.672) | 9777 (9999) | (0.339, 0.681) | 9772 (9999) | (**0.303**, 0.761) | 9214 (9000) | (0.307, **0.785**) | 9226 (9000) |
| S2 | (0.199, 0.806) | 1523 (1500) | (0.307, 0.701) | 7660 (7500) | (0.285,0.778) | 8488 (8500) | (**0.283, 0.784**) | (8578, 8500) |
| S3 | (0.164, 0.854) | 1788 (2000) | (0.156, 0.866) | 2551 (2500) | (0.220, 0.840) | 9123 (9000) | (**0.202, 0.860**) | 5605 (5500) |

Table S2. Performance measures, (NRMSE, SSIM), for CD+r applied to the raw data for all four values of μ. Bolded values indicate highest performing metric.

| R=8 | Simulation | | | | | | | |
|---|---|---|---|---|---|---|---|---|
| Subject | CD+r (μ=0.05) | | CD+r (μ=0.10) | | CD+r (μ=0.50) | | CD+r (μ=1.0) | |
| | Performance | Optimal steps | Performance | Optimal steps | Performance | Optimal steps | Performance | Optimal steps |
| S1 | (0.157, 0.868) | 1114 (1000) | (0.140, 0.887) | 1745 (1500) | (**0.136**, 0.912) | 2655 (2500) | (**0.136, 0.918**) | 2998 (3000) |
| S2 | (0.141, 0.867) | 1568 (1500) | (0.141, 0.854) | 1699 (1500) | (**0.135, 0.901**) | 2890 (3000) | (0.137, 0.884) | 2872 (3000) |
| S3 | (0.129, 0.901) | 1440 (1500) | (0.129, 0.899) | 1489 (1500) | (**0.117**, 0.933) | 4235 (4000) | (0.118, **0.934**) | 5060 (5000) |
| **R=12** | | | | | | | | |
| Subject | CD+r (μ=0.05) | | CD+r (μ=0.10) | | CD+r (μ=0.50) | | CD+r (μ=1.0) | |
| | Performance | Optimal steps | Performance | Optimal steps | Performance | Optimal steps | Performance | Optimal steps |
| S1 | (0.167, 0.865) | 1316 (1500) | (0.154, 0.875) | 1737 (1500) | (**0.148, 0.903**) | 2890 (3000) | (0.154, 0.899) | 5789 (6000) |
| S2 | (0.170, 0.826) | 1072 (1000) | (0.153, 0.864) | 1789 (2000) | (**0.147**, 0.895) | 2967 (3000) | (**0.147, 0.896**) | 3759 (4000) |
| S3 | (0.133, 0.896) | 1981 (2000) | (**0.130**, 0.916) | 2600 (2500) | (**0.130, 0.921**) | 2872 (3000) | (**0.130**, 0.903) | 2998 (3000) |
| **R=18** | | | | | | | | |
| Subject | CD+r (μ=0.05) | | CD+r (μ=0.10) | | CD+r (μ=0.50) | | CD+r (μ=1.0) | |
| | Performance | Optimal steps | Performance | Optimal steps | Performance | Optimal steps | Performance | Optimal steps |
| S1 | (0.183, 0.862) | 1391 (1500) | (0.172, 0.861) | 1675 (1500) | (**0.165, 0.889**) | 3015 (3000) | (0.193, 0.869) | 9122 (9000) |
| S2 | (0.181, 0.832) | 1333 (1500) | (0.165, 0,827) | 2060 (2000) | (**0.163**, 0.853) | 2590 (2500) | (0.164, **0.871**) | 4649 (4500) |
| S3 | (0.150, 0.857) | 1627 (1500) | (0.146, 0.878) | 1947 (2000) | (**0.143, 0.909**) | 5828 (6000) | (**0.143, 0.909**) | 6024 (6000) |
| **R=36** | | | | | | | | |
| Subject | CD+r (μ=0.05) | | CD+r (μ=0.10) | | CD+r (μ=0.50) | | CD+r (μ=1.0) | |
| | Performance | Optimal steps | Performance | Optimal steps | Performance | Optimal steps | Performance | Optimal steps |
| S1 | (0.220, 0.819) | 1699 (1500) | (**0.212**, 0.834) | 2881 (3000) | (0.240, 0.801) | 8340 (8500) | (**0.212, 0.858**) | 5959 (6000) |
| S2 | (0.233, 0.787) | 1333 (1500) | (**0.201**, 0.826) | 2881 (3000) | (0.242, 0.799) | 8299 (8500) | (0.220, 0.827) | 5898 (6000) |
| S3 | (**0.168**, 0.867) | 2890 (3000) | (0.177, 0.835) | 2239 (2000) | (0.171, **0.878**) | 6250 (6000) | (0.189, 0.864) | 8279 (8500) |

Table S3. Performance measures, (NRMSE, SSIM), for CD+r applied to the simulated data for all regularization parameter values.

| R=8 | Simulation | | | |
|---|---|---|---|---|
| Subject | L1 | LR | CD | CD+r (μ=0.10) |
| S1 | (0.262, 0.860) | (0.293, 0.802) | **(0.255, 0.873)** | (0.257, 0.865) |
| S2 | (0.271, 0.840) | (0.312, 0.752) | **(0.255, 0.861)** | (0.262, 0.846) |
| S3 | (0.239, 0.864) | (0.256, 0.832) | **(0.238, 0.869)** | (0.246, 0.853) |
| **R=12** | | | | |
| Subject | L1 | LR | CD | CD+r (μ=0.10) |
| S1 | (0.308, 0.795) | (0.336, 0.733) | (0.290, **0.835**) | (**0.284**, 0.828) |
| S2 | (0.316, 0.765) | (0.353, 0.672) | (0.285, 0.821) | (**0.277, 0.834**) |
| S3 | (0.278, 0.810) | (0.288, 0.782) | **(0.248, 0.855)** | (**0.248**, 0.852) |
| **R=18** | | | | |
| Subject | L1 | LR | CD | CD+r (μ=0.10) |
| S1 | (0.348, 0.728) | (0.369, 0.671) | **(0.311, 0.785)** | (0.313, 0.783) |
| S2 | (0.371, 0.660) | (0.393, 0.583) | (0.321, **0.766**) | (**0.316**, 0.762) |
| S3 | (0.310, 0.756) | (0.316, 0.735) | (0.274, 0.820) | **(0.268, 0.824)** |
| **R=36** | | | | |
| Subject | L1 | LR | CD | CD+r (μ=0.10) |
| S1 | (0.415, 0.591) | (0.425, 0.556) | **(0.362, 0.720)** | (0.365, 0.715) |
| S2 | (0.442, 0.485) | (0.449, 0.447) | (0.367, 0.672) | **(0.364, 0.676)** |
| S3 | (0.360, 0.660) | (0.354, 0.664) | (0.318, 0.741) | **(0.313, 0.745)** |

Table S4. Performance measures for $T_1$ mapping analysis of simulated data. Paired values in parentheses in each field correspond to the (NRMSE, CCC).